\documentclass[aps,pre,twocolumn,groupedaddress,floatfix,amsmath]{revtex4}
\usepackage{graphicx}
\usepackage{amsmath,amssymb}
\usepackage[utf8]{inputenc}
\usepackage{psfrag}

\begin{document}

\title{Curvature-driven coarsening in the two dimensional Potts model}

\author{Marcos P. O. Loureiro}
\affiliation{Instituto de F\'\i sica and INCT
Sistemas Complexos, Universidade Federal do 
Rio Grande do Sul, CP 15051, 91501-970 Porto Alegre RS, Brazil} 
\author{Jeferson J. Arenzon}
\affiliation{Instituto de F\'\i sica and INCT
Sistemas Complexos, Universidade Federal do 
Rio Grande do Sul, CP 15051, 91501-970 Porto Alegre RS, Brazil} 
\author{Leticia F. Cugliandolo}
\affiliation{Universit\'e Pierre et Marie Curie -- Paris VI, LPTHE UMR 7589,
4 Place Jussieu,  75252 Paris Cedex 05, France}
\author{Alberto Sicilia}
\affiliation{Department of Chemistry, University of Cambridge,
Lensfield Road, CB2 1EW, Cambridge, United Kingdom}

\date{\today}

\begin{abstract}
We study the geometric properties of polymixtures after a
sudden quench in temperature. We mimic these systems with
the $q$-states Potts model on a square lattice with and without 
weak quenched disorder, and their evolution
with Monte Carlo simulations with non-conserved
order parameter. We analyze the distribution of hull 
enclosed areas for different initial conditions and 
compare our results with recent exact and numerical findings
for $q=2$ (Ising) case. Our results demonstrate the memory 
of the presence or absence of long-range correlations in the initial 
state during the coarsening regime and exhibit super-universality 
properties.
\end{abstract}

\maketitle

\section{Introduction}

Domains formed during the evolution of mixtures are of both
theoretical and technological importance, applications including
foams~\cite{GlAnGr90}, cellular tissues~\cite{MoAlIg93},
superconductors~\cite{PrFiHoCa08}, magnetic
domains~\cite{BaSeWe90,Jagla04}, adsorbed atoms on surfaces, {\it
etc.}  In particular, in metallurgy and surface science, the
polycrystalline microstructure, and its time evolution, are important
in determining the material properties.

After a sudden change in temperature (or in another suitable control
parameter) that takes the system from the high temperature phase into
the coexistence region, the system tends to organize in progressively
larger ordered structures.  The system's temporal evolution is ruled
by thermal, diffusive, and curvature driven processes, and the actual growth law
depends on general features suchlike the presence of quenched
disorder, the dimension of the order parameter and whether it is
conserved or not. Although much is known for binary mixtures and
systems with twofold ground degeneracy ($q=2$), much less is
understood for polymixtures and manifolded ground states ($q>2$). In
the latter case, topological defects pin the domain wall dynamics and
even in pure systems thermal activation is necessary to overcome the corresponding energy
barriers.

Many interesting cellular growth processes are captured by a
curvature-driven ordering processes in which thermal effects play a
minor role. These are ruled by the Allen-Cahn equation in which the
local velocity of an interface is proportional to the local curvature,
$v=-(\lambda/2\pi) \kappa$, where $\lambda$ is a temperature and
$q$-dependent dimensional constant related with the surface tension
and mobility of a domain wall and $\kappa$ is the local curvature. The
sign is such that the domain wall curvature is diminished along the
evolution.  In $d=2$ the time dependence of the area contained within
any finite domain interface (the hull) on a flat surface is obtained
by integrating the velocity around the hull and using the Gauss-Bonnet
theorem:
\begin{equation}
\frac{dA}{dt}= \oint v dl = -\frac{\lambda}{2\pi}\oint \kappa dl
=-\lambda\left(1-\frac{1}{2\pi}\sum_i \alpha_i\right),
\end{equation}
where $\alpha_i$ are the turning angles of the tangent vector to the
surface at the $n$ possible vertices or triple junctions.  In some
systems, such as the Ising $q=2$ model, $\sum_i \alpha_i=0$ since
there are no such vertices and we obtain $dA/dt=-\lambda$ for all
hull-enclosed areas, irrespective of their
size~\cite{Mullins56,ArBrCuSi07}. In highly
anisotropic systems, as the Potts model studied here, the angles are
all different. In some of the systems that motivate this
work, like soap froths, the angles are all equal to 
$2\pi/3$, that is, $\alpha_i=\pi/3, \forall i$. 
Focusing on systems in which there is a one-to-one correspondence 
between vertices and sides (meaning that we exclude the Ising limit
in which there are no vertices but each wall has one side) 
the above equation reduces to the von
Neumann law~\cite{Neumann52,Mullins56} for the area $A_n$ of an
$n$-sided hull-enclosed area:
\begin{equation}
\frac{dA_n}{dt}=\frac{\lambda}{6}(n-6) \qquad\qquad n>1.
\label{eq:Neumann}
\end{equation}
Whether a cell grows, shrinks or remains with constant area depends
on its number of sides being, respectively, larger than, 
smaller than or equal to 6.
The above law can be extended to the case in which the typical internal
angle depends on the number of sides~\cite{GlSt89} and to 
$3d$~\cite{MaSr07} as well.

Potts models, that is to say $q$-state spin models on a lattice,
simulate grain growth.  Early studies of the $2d$ $q$-state Potts
model with non-conserved order-parameter dynamics have shown that the
growth law is analogous to that of the (non conserved) Ising model,
where in the scaling regime the characteristic length scale follows
the $t^{1/2}$ Allen-Cahn law~\cite{GrAnSr88,LaDaVa88}, regardless of
the value of $q$.  However, for $q>2$ and $T=0$, this power law growth
does not hold for long times, the characteristic length scale
converging to a limiting value. This is in agreement with the
Lifshitz-Safran criterium~\cite{Lifshitz62,Safran81} that states that
when the ground state degeneracy is $q\geq d+1$, where $d$ is the
system dimensionality, there might be domain wall pinning depending on
the lattice geometry and the density of topological defects.  In the
$2d$ Potts model~\cite{Wu82} with $q\geq 3$ such point defects are the
convergence of three distinct phases
borders~\cite{Lifshitz62,Safran81}. The asymptotic density of defects
that depends on $q$ (negligible for $q\leq 4$ but nonzero for $q>4$)
was also related with glassy behavior~\cite{OlPeTo04,FeCa07}.  In the
presence of thermal fluctuations, the excess energy due to too many domain
borders disappears after a transient~\cite{BeFeCaLoPe07}.

In a sequence of papers~\cite{ArBrCuSi07,SiArBrCu07,SiSaArNrCu09}, 
we explored the
time evolution of domain and hull-enclosed area distributions in
bidimensional coarsening systems with scalar order parameter. To study the 
non-conserved order parameter case, on the
one hand we used a continuum description based on the Ginzburg-Landau
equation for the scalar field and the Allen-Cahn law $dA/dt=-\lambda$.  
On the other hand, we studied the
kinetic Ising model on a square lattice.  For the $q=2$ Ising case
with non-conserved order parameter, the absence of triple points leads
to the uniform shrinkage of all hull-enclosed areas. More concretely,
the number of hull-enclosed areas per unit system area, 
in the interval
$(A,A+dA)$ at time $t$ is related to the distribution at the initial
time $t_i$ through
\begin{equation}
n_h(A,t) = n_h\left[ A + \lambda(t-t_i),t_i \right]
\; . 
\label{eq.map}
\end{equation}
Once the initial distribution is known, the above
equation gives the distribution at any time $t$. For example, if 
one takes as initial states at $t_i=0$ configurations in 
equilibrium state at $T_c$, where the
distribution is exactly known for $q=2$~\cite{CaZi03},
$n_h(A,0)=c_h/A^2$ with $c_h=8\pi\sqrt{3}$, the distribution at $t>0$
is $n_h(A,t)=c_h/(A+\lambda_h t)^2$, that compares extremely well with
simulation data for the Ising model on a square
lattice~\cite{ArBrCuSi07}. For a quench from infinite temperature, on
the other hand, the initial distribution corresponds to the one of the
critical random continuous
percolation~\cite{ArBrCuSi07,SiArBrCu07}. This observation is an
essential ingredient to understand the fact that, and obtain the
probability with which, the system attains a striped frozen state at
zero temperature (see~\cite{BaKrRe09} and references therein).
Interestingly, on a square lattice, the initial state does not
correspond to the random percolation critical point but the coarsening
evolution gets very close to it after one or two Monte Carlo
steps. Therefore, although Eq.~(\ref{eq.map}) was obtained with the
continuous description and, a priori, it is not guaranteed to apply on
a lattice, it does describe the coarsening dynamics of the discrete
Ising model remarkably accurately.
In the presence of vertices, instead, the single hull-enclosed area
evolution depends on $n$ and, notably, a given hull-enclosed area can
either shrink or grow depending on whether $n<6$ or $n>6$, respectively. 
Therefore, one cannot write a simple relation as the one in
Eq.~(\ref{eq.map}) to link the area distribution at time $t$ to the
one at the initial time $t_i$ and the distribution might get
scrambled in a non-trivial way during the coarsening process
(for example, when a domain disappears, the number of sides of the 
neighboring domains changes, along with their growth rate).  

In this paper we study the distribution of hull-enclosed areas
during evolution in a Potts model with different number of states,
notably $q\leq 4$ and $q>4$. We then 
try to give an answer to some questions that can be posed when
dealing with more than two competing ground states. To what extent the
results obtained for $q=2$ are also valid for $q\geq 3$?  In particular, in
a quench from infinite temperature, does the random percolation
critical point affect the evolution as it does for $q=2$?  Moreover,
which is the interplay between nucleation and growth when the system
goes through a first-order phase transition as in the cases $q>4$?
What happens when the starting point of the quench is a first order
transition point, with finite range correlations? In the presence of
weak disorder, when first order transitions become continuous, are the
scaling functions universal (super-universality)?  

The paper is organized as follows. In section~\ref{section.Potts}
we introduce the Potts model
and, in section \ref{section.eq-time.corr}, we study the space-time correlation 
function starting from
differently correlated initial states. Then, in section
\ref{section.nA}, several area distribution functions, either in the
initial state or after the quench in temperature, are studied and
described in detail. Afterwards, in section \ref{section.disorder},
the Potts model with ferromagnetic random bonds (weak disorder) is
studied and compared with the pure model. Finally, we make some final
remarks and conclude.

\section{The Potts Model}
\label{section.Potts}

We consider the Potts model~\cite{Wu82} on a square lattice of
$N=L^2$ spins with $L=10^3-5\times 10^3$, 
the state variables of which, $s_i$ with 
$i=1\ldots N$, assume integer values from 1 to $q$. The Hamiltonian is given by
\begin{equation}
{\cal H} = -\sum_{\langle ij\rangle} J_{ij}\delta_{s_is_j}
\end{equation}
where the sum is over nearest-neighbor spins on the lattice and,
until section \ref{section.disorder}, $J_{ij}=1,\;\forall i,j$.  The
transition, discontinuous for $q>4$ and continuous for $q\leq 4$,
occurs at $T_c=2/\ln (1+\sqrt{q})$. We run from 500 to 4000 Swendsen-Wang 
(SW) algorithm steps to reach an equilibrium initial condition at
the critical point, and average over 1000
samples to build the correlations and distributions. 
After equilibrium is attained and
the system quenched, the evolution follows the heat-bath Monte Carlo 
algorithm~\cite{NeBa99} and all times are given in Monte Carlo step (MCs)
units, each one corresponding to a sweep over $N$ randomly chosen spins.

\begin{figure}[h]
\psfrag{r}{$r/R(t)$} \psfrag{cr}{$C(r/R(t))$} \psfrag{t}{$t$}
\psfrag{r2}{$R^2(t)$} \psfrag{tt}{$t$}
\includegraphics[width=9cm]{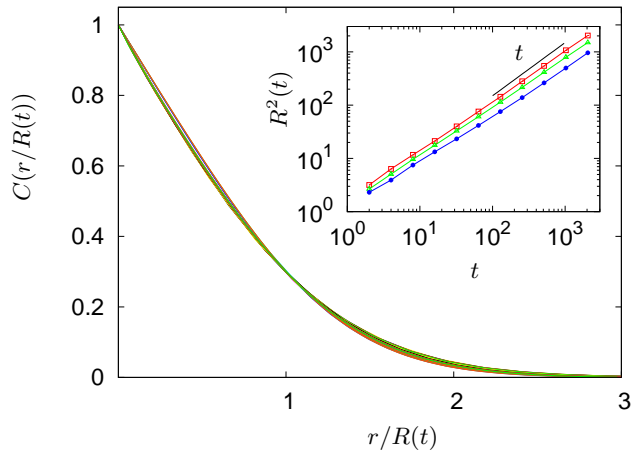}
\caption{(Color online) Rescaled space-time correlation function at
  several times (from $t=2^4$ to $2^{11}$ MCs) after a quench from
  $T_0\to\infty$ to $T_f= T_c/2$ for $q=2$, 3 and 8, with and without
  weak disorder (the disordered case is discussed in section
  \ref{section.disorder}).  A good collapse is obtained for all $q$
  when rescaling the spatial variable by $R(t)$ obtained from
  $C(R,t)=0.3$ (note that the working temperature is different in all
  cases).  Inset: $R^2(t)$ against $t$ in a double logarithmic scale
  for $q=2$, 3 and 8 (from top to bottom).  The characteristic length,
  related to the average domain radius, depends weakly on $q$ and $T$
  for the pure model (through the prefactor). Within the time window
  explored there are still some small deviations from the expected
  Allen-Cahn behavior, $R(t) \simeq t^{1/2}$ in the case $q=8$.  When
  weak disorder is introduced (not shown), the growth rate is greatly
  reduced and $R^2$ deviates from the linear behavior.}
\label{fig.crtinf}
\end{figure}

After a quench in temperature, the abrupt increase in the local
correlation creates an increasing order, despite the competition
between different coexisting stable phases. In the cases $q\leq 4$ the
paramagnetic state becomes unstable at $T_c$ and after a quench to any
non-zero sub-critical temperature the system orders locally and
progressively in patches of each of the equilibrium phases; in other
words, it undergoes domain growth. In the cases $q>4$ one can
distinguish three working temperature regimes: at very low $T_f$ the
system gets easily pinned; at higher $T_f$ but below the spinodal
temperature $T_s$ at which the paramagnetic solution becomes unstable
the system undergoes domain growth; above the spinodal $T_s$ there is
competition between nucleation and growth and coarsening. However, the
latter regime is very hard to access since $T_s$ is very close to
$T_c$ (e.g. $T_s\simeq 0.95 T_c$ for $q=96$~\cite{LoFrGrCa09}). In
this work we focus on the dynamics in the intermediate coarsening
regime.

The detailed evolution of the system during the coarsening
dynamics depends on the correlations already present in the initial
state, whether absent ($T_0\to\infty$), short- ($T_c<T_0<\infty$ for all
$q$ and also $T_0=T_c$ for $q>4$) or long-range ($T_0=T_c$ for $q\leq 4$). In
the latter case there is already one spanning cluster at $t=0$, since
the thermodynamic transition also corresponds to a percolation
transition in $2d$. On the other hand, for short-range initial
correlations, such spanning domains are either formed very fast
(e.g. in the case $q=2$ with $T_0\to\infty$), 
or not formed at all (or at least not within the timescales
considered here).  

\section{Equal-time correlations}
\label{section.eq-time.corr}

The degree of correlation between spins is measured
by the equal time correlation function
\begin{equation}
C(r,t) = \frac{q}{q-1}\left( \left\langle\delta_{s_i(t)s_j(t)}
\right\rangle_{|i-j|=r}  - \frac{1}{q}\right),
\end{equation}
where the average is over all pairs of sites a distance $r$ apart.
Away from the critical temperature, correlations are short-range,
and after a quench from $T_0>T_c$ to $T_f<T_c$ these initial correlations
become irrelevant after a finite time and the system looses memory of the
initial state. In this sense, equilibrium states at all temperatures
above $T_c$ are equivalent. 

Figure~\ref{fig.crtinf} shows the correlation function $C(r,t)$ as a
function of the rescaled distance, $r/R(t)$, after a quench from
infinite temperature, where the initial correlation is null, to a
working temperature $T_f=T_c/2$ for $q=2$, 3 and 8.  In the inset, we
show the length scale $R(t)$ computed as the distance $r$ at which the
correlation has decayed to $0.3$ of its initial value, that is,
$C(R,t)=0.3$. The linear behavior $R^2\sim t$ is clearly seen for
$q=2$.  For $q>2$ there are still some deviations from the $t^{1/2}$
law at early times but these disappear at longer times (see
also~\cite{GrAnSr88,LaDaVa88}).  Deviations at short times are indeed
expected, due to pinning at low temperatures (note that $T_f=T_c/2$
decreases with $q$).  An excellent collapse of the spatial correlation
is observed, as expected from the dynamic scaling hypothesis, when $r$
is rescaled by the length $R(t)$. Moreover, the universal curves seem
to coincide for these values of $q$, regardless of the different
temperatures after the quench, in accordance with previous
evidence~\cite{KaNiGu85,KuGuKa87,LaDaVa88} for the fact that the
space-time correlation scaling function (and its Fourier transform,
the structure factor) are insensitive to the details of the underlying
Hamiltonian.  Whether this apparent $q$ independence is only
approximate or exact, and in the latter case whether it remains valid
for very large values of $q$, are still open
issues~\cite{LiMa93,RaMa05}.  As for the Ising model~\cite{HuBr92},
the small $r$ behavior is linear for all $q$, in agreement with
Porod's law~\cite{Bray94,LiMa93}.

\begin{figure}[h]
\psfrag{r}{$r/R(t)$}
\psfrag{cr}{$C(r,t)$}
\psfrag{t}{$r$}
\psfrag{t4}{$t=4$}
\psfrag{t10}{$t=2^{10}$}
\psfrag{r2}{$C(r,t)$}
\psfrag{r4}{$r^{-1/4}$}
\psfrag{tt}{$t^{1/2}$}
\includegraphics[width=9cm]{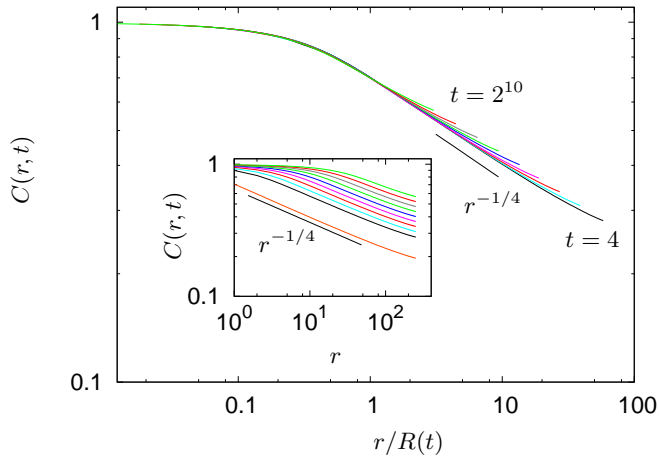}
\caption{(Color online)
Rescaled space-time correlation function at several times
($t=4,\ldots,2^{10}$ MCs) after a quench from $T_0=T_c$ to 
$T_f=T_c/2$ for the Ising $q=2$ case in log-log scale
(cfr. Figs.~\ref{fig.crtc_q3} and~\ref{fig.crtc_q8} for $q=3$ and 
$q=8$, respectively).  A very
good collapse is obtained when rescaling the spatial variable by
$R(t)$, obtained from $C(R,t)=0.7$.  At $t=0$, the correlation decays
as $r^{-\eta}$ ($\eta=1/4$), while for $t>0$ partial memory of the
initial state is preserved, as evidenced by the power-law tail. 
The inset presents $C(r,t)$ for the same
times used in the main figure, with no rescaling of $r$. The
correlation at $t=0$ is also shown, and deviations from the power-law
behavior are already present, due to the strong fluctuations in
magnetization at $T_c$.
}
\label{fig.crtc}
\end{figure}

\begin{figure}[h]
\psfrag{r}{$r/R(t)$}
\psfrag{cr}{$C(r,t)$}
\psfrag{t}{$t$}
\psfrag{r2}{$R^2(t)$}
\psfrag{r4}{$r^{-4/15}$}
\psfrag{tt}{$t$}
\includegraphics[width=9cm]{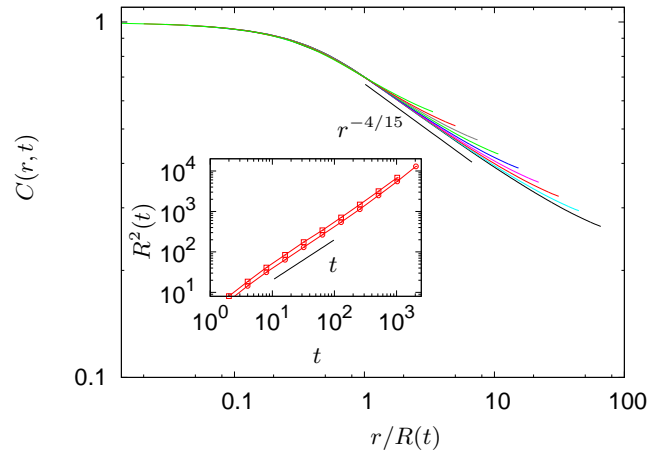}
\caption{(Color online) Rescaled space-time correlation function at
  several times ($t=4,\ldots,2^{10}$ MCs) after a quench from $T_0=T_c$ to
  $T_f=T_c/2$ for $q=3$ in log-log scale (cfr. Figs.~\ref{fig.crtc}
  and~\ref{fig.crtc_q8} for $q=2$ and $q=8$, respectively).  The slope
  $r^{-\eta}$ with $\eta=4/15$ is shown as a guide to the eye. 
  Inset:  $R^2(t)$ for $q=2$ (top) and 3 (bottom), obtained from
$C(R,t)=0.7$ together with the law  $R^2\simeq t$.}
\label{fig.crtc_q3}
\end{figure}
 
On the other hand, when the starting point is an equilibrium state at
$T_c$ with long-range correlations ($q\leq 4$)  obtained by performing a sufficient 
number of Swendsen-Wang
steps~\cite{NeBa99}, the system keeps memory of the long-range
correlations present in the critical initial state.  The
equal time equilibrium correlation function decays at the critical
temperature as a power-law, $C(r,0)\sim r^{2-d-\eta}$, where both the
coefficient and the exponent $\eta$ depend on $q$. For example, in two
dimensions, $\eta=1/4$ for $q=2$ and $\eta=4/15$ for
$q=3$~\cite{Wu82}.  
Figure~\ref{fig.crtc} shows the
rescaled correlation function at several instants after the quench for
$q=2$. Some
remarks are in order.  First, there is a very good collapse for length
scales up to $r\sim R(t)$, where $R(t)$ is such that $C(R,t)=0.7$.
Second, deviations from the initial power-law occur both at short and
long length scales. Due to the ever growing structures, correlations
decrease very slowly for small $r$.  Indeed, $C$ almost follows the
plateau close to unity up to a certain, time increasing, distance. For
long distances, on the other hand, these deviations occur because the
initial states can be magnetized (at $T_c$, magnetization goes as
$L^{-\beta/\nu}$) and, as $r$ increases and the spins decorrelate,
$C(r,t)$ attains a plateau at $m^2(t)$. Since,
eventually, the system equilibrates, the longer the time, the larger
the magnetization and, consequently, the higher the plateau.
Although not done here, it is also possible to postpone the approach
to equilibrium by choosing initial states with very small
magnetization, as done by Humayun and Bray~\cite{HuBr91}, who also
introduced a correction factor to account for boundary effects due to
the system size being much smaller than the correlation length,
$L\ll\xi$, at $T_c$. Essentially the same behavior is obtained for
$q=3$  with $\eta=4/15$  (shown in Fig.~\ref{fig.crtc_q3})
and $q=4$ with $\eta=1/2$ (not shown), respectively.
We obtain
  the same scaling function for different final temperatures (not shown),
  indicating that super-scaling holds with relation to temperature.

\begin{figure}[h]
\psfrag{r}{$r/R(t)$}
\psfrag{cr}{$C(r,t)$}
\psfrag{tt}{$t$}
\psfrag{r2}{$R^2(t)$}
\psfrag{t}{$t$}
\includegraphics[width=9cm]{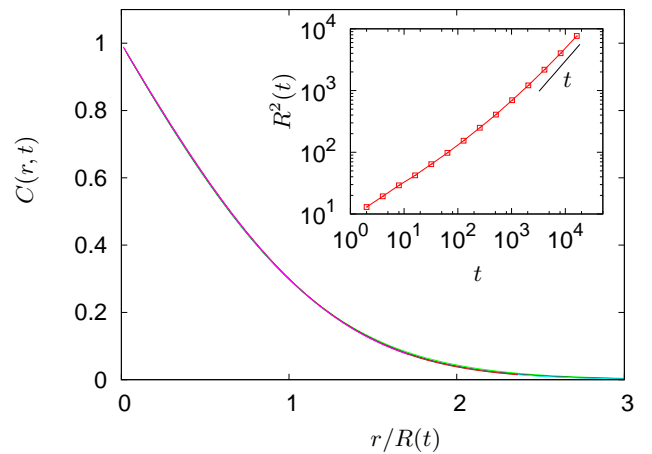}
\caption{(Color online) Rescaled space-time correlation function at
  several times ($t=2^{10},\ldots,2^{14}$ MCs) after a quench from
  $T_0=T_c$ to $T_f=T_c/2$ for $q=8$ in linear-linear scale
  (cfr. Figs.~\ref{fig.crtc} and~\ref{fig.crtc_q3} for $q=2$ and
  $q=3$, respectively). }
\label{fig.crtc_q8}
\end{figure}

For $q>4$, differently from the previous cases, there are no
long-range correlations at $T=T_c$, only finite-range
ones. The actual correlation length at $T_c$ can be obtained
analytically~\cite{BuWa93}, but we did not attempt to measure it,
since the correlator that we use does not remove power-like
prefactors~\cite{JaKa95}.  For quenches below the
limit of stability  of the paramagnetic high-temperature state, 
see Fig.~\ref{fig.crtc_q8} for $q=8$, the finite correlation length at $t=0$
 is washed out once the scaling regime is
attained. As  might have been expected, the scaling function is
  indistinguishable, within our numerical accuracy, from the one for
  systems with a continuous phase transition ($q=2,\ 3$) quenched from
  infinite temperature, cfr. Fig.~\ref{fig.crtinf}.
The collapse is not as good at short times
  since it takes longer for the system to approach the scaling regime
  in the $q=8$ case (see the inset).  

After some discussion it is now well-established that weak quenched
disorder changes the order of the phase transition from first-order to
second-order when $q>4$. The critical exponent $\eta$ depends on $q$
although very weakly. The scaling function of the space-time
correlation function after a quench from the critical point also
depends on $q$~\cite{JaPi00,AuIg03}.

\section{Area Distributions}
\label{section.nA}

More insight on the growing correlation observed during the coarsening
dynamics after the quench can be gained from the study of the
different area distributions $n(A,t)$. We consider two measurements:
geometric domains and hull-enclosed areas. Geometric domains are
defined as the set of contiguous spins in the same state. In
principle, such domains may enclose smaller ones that, in turn,
may also enclose others and so on.  One can also consider the external
perimeter of the geometric domains (the hull) and the whole area
enclosed by it. Let us discuss the distribution of these areas after a
quench from higher to low temperatures.

\subsection{The initial states}

Prior to the quench, the system is prepared in an initial state having
either zero, finite or infinite range correlation, corresponding to
$T_0=\infty$ ($\forall q$), $T_0=T_c$ ($q>4$) and $T_0=T_c$ ($q\leq
4$), respectively.  Infinite temperature states are created by
randomly assigning, to each spin, a value between 1 and $q$, while to
obtain an equilibrium state at $T_c$, the system is further let evolve
during a sufficient number of Swendsen-Wang Monte Carlo steps at $T_c$
(typically 500).

\begin{figure}[h]
\psfrag{A}{$A$}
\psfrag{log}{$\ln [A \ n_{h,d}(A,0)]$}
\psfrag{8}{8}
\psfrag{5}{5}
\psfrag{4}{4}
\psfrag{q=3}{$q=3$}
\psfrag{logAn}{$\log An(A,0)$}
\includegraphics[width=8cm]{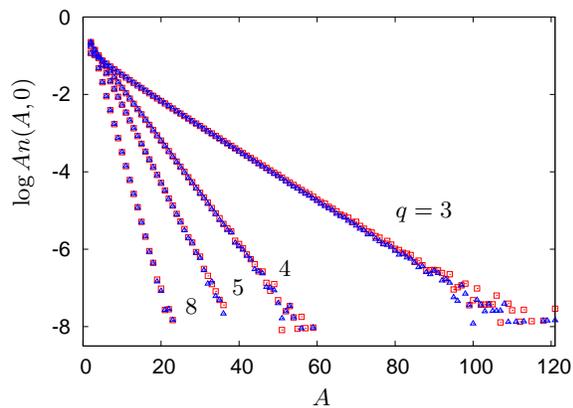}
\caption{(Color online) Equilibrium domain (blue triangles) and
  hull-enclosed area (red squares) distributions at $T_0\to\infty$ for
  several values of $q$, corresponding to the random percolation
  problem at $1/q$ occupation density. The data are shown in
  linear-log scale. }
\label{fig.dist_2d_t0}
\end{figure}

\subsubsection{$T_0\to\infty$}

When the temperature is infinite, neighboring spins are uncorrelated,
and the configurations can be mapped onto those of the random
percolation model with equal occupation probability $p=1/q$.  On the
square lattice considered here, this state is not critical for all
values of $q$ (criticality would require that one of the species
density be $0.59$). The highest concentration of a single species,
$0.5$, is obtained for $q=2$. This corresponds to the continuous
random percolation threshold but on a lattice, however, it is
critical percolation on the triangular case only. Still, for the $q=2$
Ising model on a square lattice, as shown in \cite{ArBrCuSi07}, the
proximity with the percolation critical point strongly affects the
system's evolution.  

Figure~\ref{fig.dist_2d_t0} shows the geometric
domains and hull-enclosed area distributions
$n(A,0)=qn_{\scriptstyle\rm rp}(A,1/q)$, where $n_{\scriptstyle\rm
  rp}(A,p)$ is the corresponding size distribution for random
percolation with occupation probability $p=1/q$. The factor $q$ comes
from the fact that the $q$ species equally contribute to the
distribution, while in the percolation problem, a fraction $1-1/q$ of
the sites is empty.  Unfortunately, $n_{\scriptstyle\rm rp}$ is not
known analytically for general values of $p$ away from $p_c$.  
For $q=2$ (not shown, see~\cite{SiArBrCu07}), the
  system is close to the critical percolation point, and the
  distributions are power-law, $n(A,0) \simeq A^{-\tau}$ with $\tau=2$
  for hull-enclosed areas and $\tau\gtrsim 2$ for domain
  areas, up to a large area cut-off where they cross over to
  exponential decays.  For $q=3$, deviations from linearity are
  already perceptible and, moreover, the hull-enclosed and domain area
  distributions can be resolved. For
sufficiently low $p$ (but probably valid for all $p<p_c$) the
distribution tail is
exponential~\cite{MuSt76,KuSo78a,Schwartz78,StDo79,Hoshen79,ScBr85}
and the simulation data is compatible with~\cite{PaSo81,StAh94}
$n_{\scriptstyle\rm rp}(A,p)\sim A^{-\theta}\exp [-f(p)A]$, where the
exponent $\theta$ depends on the dimension only ($\theta=1$ for
$d=2$).  Since all $q$ states are equally present, the domains are
typically smaller the larger the value of $q$ and the occupation
dependent coefficient $f(p)$ thus increases for increasing $q$ or
decreasing $p$.  The data for the different distributions shown in
Fig.~\ref{fig.dist_2d_t0} can only be resolved for large values of $A$
and small values of $q$. These differences are due to domains embedded
into larger ones, already present in the initial state, that may
remain at long times after the quench. For larger values of $q$ the
  distributions get closer to an exponential in their full range of
  variation and it becomes harder and harder to distinguish
  hull-enclosed and domain area probability distribution
functions (pdfs).

\begin{figure}[h]
\psfrag{A}{$A$}
\psfrag{nh}{$\displaystyle \frac{n_h(A,0)}{q-1}$}
\psfrag{nh2}{\hspace{-4mm}$\small \frac{n_h(A,0)A^2}{q-1}$}
\psfrag{q10}{$10$}
\psfrag{q8}{$q=8$}
\includegraphics[width=8cm]{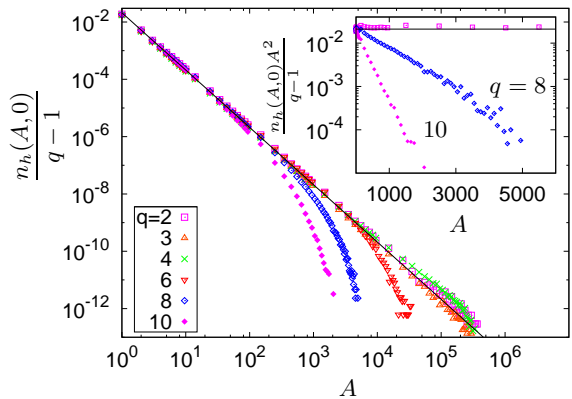}\hspace{-1mm}
\caption{(Color online)
Equilibrium power-law distribution of hull-enclosed areas at $T_c$ for
several values of $q$, from 2 to 8. The line is $c_h^{(2)}/A^2$, from 
Eq.~(\ref{eq.tct0}). 
Taking $c_h^{(q)}$ as a fitting parameter one finds very small values 
obeying $c_h^{(4)}<c_h^{(3)}<c_h^{(2)}$ (see text). 
For $q>4$, the transition is discontinuous but nonetheless the distribution
follows the power law until a crossover length, beyond which it 
decays exponentially, as shown in the inset for $q=8, \ 10$
and, for comparison, $q=2$ (the horizontal data).}
\label{fig.tct0}
\end{figure}

\subsubsection{$T_0=T_c$ and $2\leq q\leq 4$}

When the transition is second order and the system is equilibrated 
at the critical temperature $T_c$, the distribution of both domains
and hull-enclosed areas follow a power law. For example, for $q=2$ 
in $d=2$, the hull-enclosed area distribution is given by
$n_h(A,0)=c_h^{(2)}/A^2$, where
$c_h^{(2)}=1/8\pi\sqrt{3}\simeq 0.0229$~\cite{CaZi03}. Generally, 
 the hull-enclosed area distribution
for $q=2$, 3 and 4, is found to be 
\begin{equation}
n_h(A,0) = \frac{(q-1) c_h^{(q)}}{A^2}
\label{eq.tct0}
\end{equation}
as can be seen in Fig.~\ref{fig.tct0}. We choose to use the $q-1$
prefactor instead of $q$ for consistency with our previous
work. This can be done, however, with a small modification of the
constant, $(q-1)c_h=qc_h'$. Thus, each spin species contributes with
the same share to the total distribution. Notice that, unless
for $q=2$, the value of $c_h^{(q)}$ is not
known exactly.  A rough estimate of the constants can be obtained by
taking them as fit parameters; we obtain
$c_h^{(3)}\simeq 0.0203$ and $c_h^{(4)}\simeq 0.0192$. The fit for
$q=2$ yields $c_h^{(2)}\simeq 0.0227$, that compares well with the
exact value.  The constants obey the inequality
$c_h^{(4)}<c_h^{(3)}<c_h^{(2)}$.  

For geometric domains, a relation similar to Eq.~(\ref{eq.tct0}) is
obtained for the area distribution at $T_c$ and $2\leq q\leq 4$.  The
exponent is slightly larger than 2,
$\tau=379/187$ for $q=2$~\cite{StVa89,StAh94,SiArBrCu07}, and seems to 
be independent of $q$ for $q=3,\ 4$. The coefficients, however, are not
exactly known (numerically, for $q=2$, it is close to
$c_h^{(2)}$~\cite{SiArBrCu07}).

\subsubsection{$T_0=T_c$ and $q>4$}

When the transition is discontinuous, the power law 
exists up to a crossover length, typically of the order of the
correlation length $\xi$, where the distribution deviates and falls off
faster.  Depending on the values of $\xi$ and
$L$, the crossover may or may not be observed 
due to the weakness of the transition. 
For example, for $q=5$, 6 and 8,
estimates~\cite{BuWa93,DeYo02} of $\xi$ are, 2512, 159 and 24,
respectively. For $q=5$, $\xi$ is indeed larger than the system size 
$L$ that we are using and
the system behaves as it were critical. Figure~\ref{fig.tct0} also presents data
for the hull-enclosed area distributions in models with $q=6$, 8 and 10, where 
the deviations from the power-law are very
clear, and occurring at smaller values of $A$ for increasing $q$, as
expected. Equation~(\ref{eq.tct0}) is thus valid up to this crossover
length. 
The inset of Fig.~\ref{fig.tct0} shows that for areas that are larger
than a certain value $A^*_q$, the distribution decays exponentially
for $q>4$.  Thus, the general form of the hull-enclosed area
distribution, apparently valid for all values of $q$ at $T_c$, is
\begin{equation}
n_h(A,0) = \frac{(q-1) c_h^{(q)}}{A^2} e^{- \alpha_q
					    \frac{A}{A^*_q}}
\end{equation}
where $\alpha_q=0$ for $q\leq 4$. 

The area distribution of geometric domains is very similar to the
hull-enclosed area one. None of the coefficients $c_d^{(2)}$, are
exactly known; the numerical results for $q=2$ described
in~\cite{SiArBrCu07} suggest that $c_d^{(q)}$ is very close to
$c_h^{(2)}$ and the study of the distributions for $q=3,\ 4$ suggests
that the similarity between $c_d^{(q)}$ and $c_h^{(q)}$ holds for
general $q$ as well.

\subsection{The coarsening regime}

\begin{figure}[h]
\includegraphics[width=4.2cm]{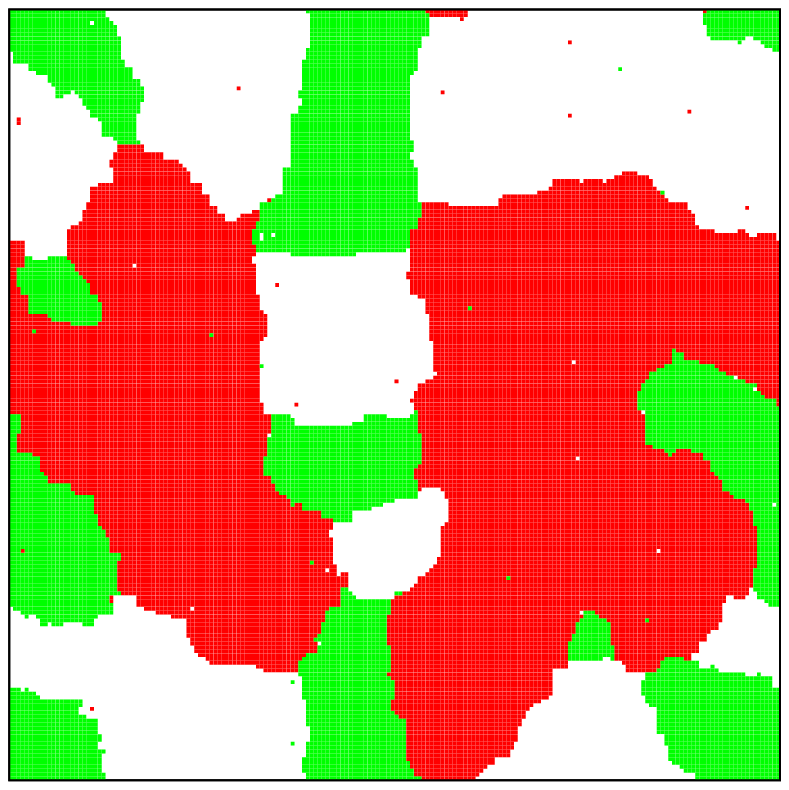}\hspace{-1mm}
\includegraphics[width=4.2cm]{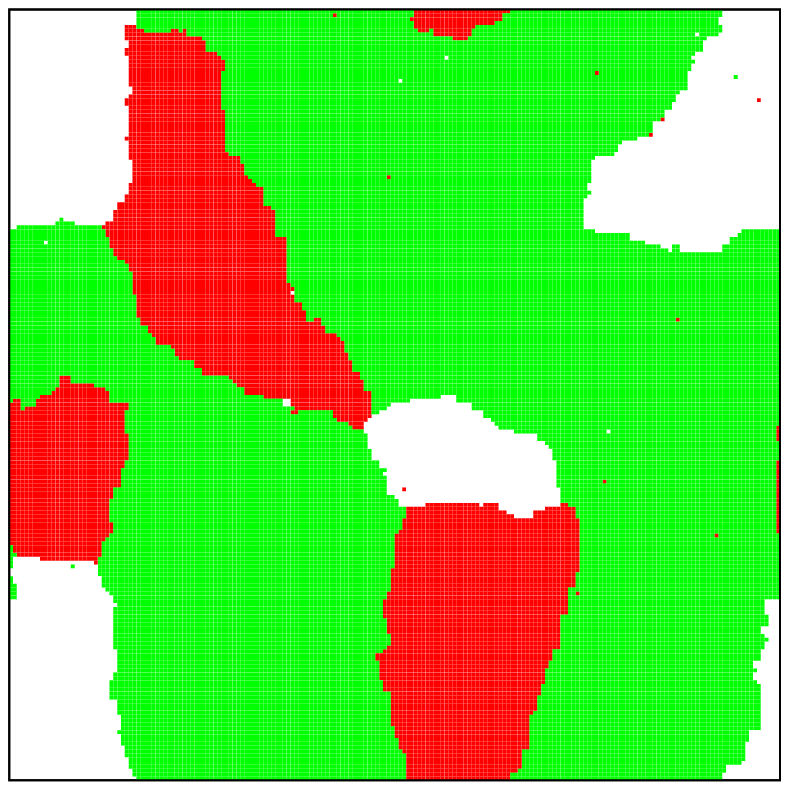}\hspace{-1mm}
\centerline{(a)\hspace{4cm}(b)}
\includegraphics[width=4.2cm]{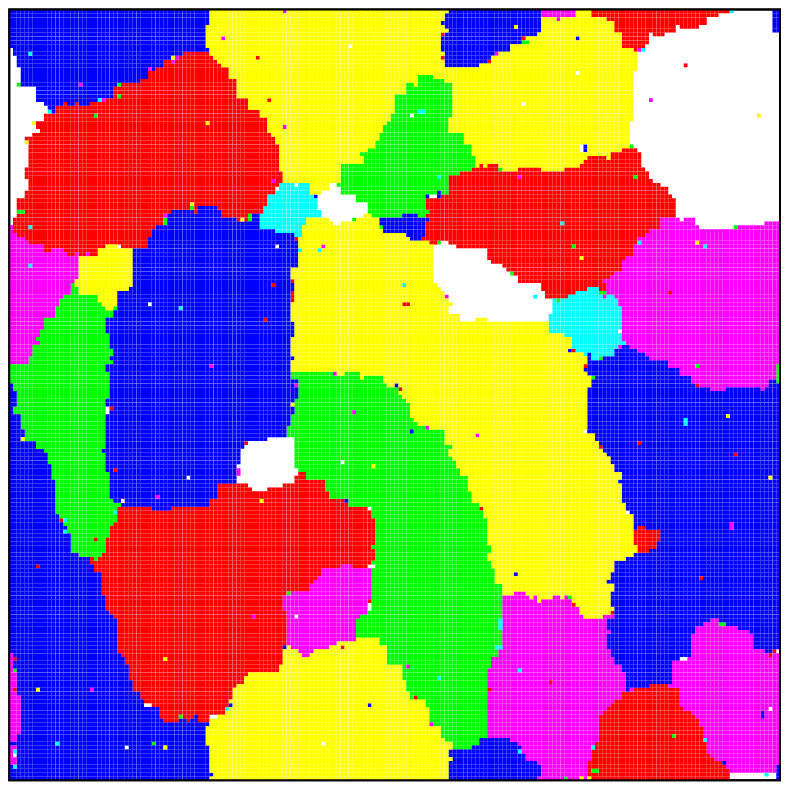}
\includegraphics[width=4.2cm]{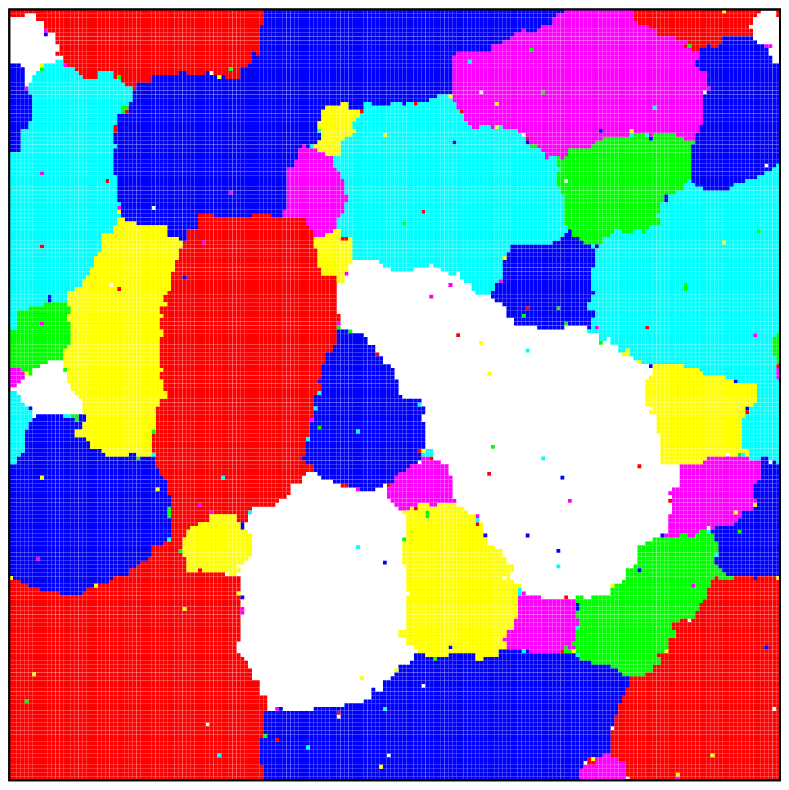}
\centerline{(c)\hspace{4cm}(d)}
\caption{(Color online) Snapshots at $t=2^{10}$ MCs after a quench
from $T_0\to\infty$ and $T_0=T_c$, left and right, respectively to
$T_f=T_c/2$, for $q=3$ [(a) and (b)] and $q=7$ [(c) and
(d)]. Different colors correspond to different species. Notice
that, for $q=7$, some domains have just coalesced (for example, the
large yellow domain in panel (c), an effect that is absent for $q=2$ and
becomes rarer as $q$ increases. Figure~\ref{fig.snapshot4} gives a more
detailed view of such occurrence. A few thermal fluctuations are also
visible as small dots inside the clusters.  Note that the structures
are typically larger on the right column ($T_0=T_c$) than on the left
one ($T_0\to\infty$) for $q=3$. This is not the case for $q=7$
where there are no big structures in the initial state since
$\xi(q=7,T_c)$ is finite.}
\label{fig.snapshot3}
\end{figure}

\begin{figure}[h]
\psfrag{t0}{t=500}
\psfrag{t1}{600}
\psfrag{t2}{700}
\psfrag{t3}{800}
\includegraphics[width=9.cm]{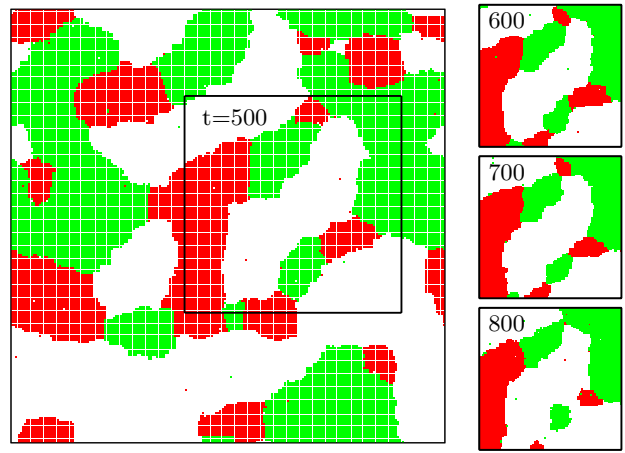}
\caption{(Color online) Snapshots at several times after a quench
from $T_0\to\infty$ to $T_f=T_c/2$, for $q=3$.
Different colors correspond to the 
three different species. The snapshots show the coalescence 
of two white domains
inside the $100\times 100$ zoomed region. Notice also
the rare occurrence of a domain fully embedded into a single
domain at $t=800$ MCs. }
\label{fig.snapshot4}
\end{figure}

In Fig.~\ref{fig.snapshot3} we show snapshots of the system at
$t=2^{10}$ MCs after the quench for different values of $q$
and $T_0$. When the initial condition is equilibrium at $T_c$ for
$2\leq q\leq 4$, the system presents larger domains than when the 
initial conditions are random, since the latter has an exponential 
instead of a power-law size distribution. On the other hand, for
$q>4$, when the correlation length at $T_c$ is finite, there is
no visible difference in the snapshots. In these figures, with the 
exception of some very small thermally induced fluctuations, there is no
domain fully embedded inside another one, at variance with 
the $q=2$ behavior~\cite{SiArBrCu07}. Nonetheless, in 
Fig.~\ref{fig.snapshot4}, one such rare embedded domain occurs, being 
created after a coalescence process among the neighbors. 
Thus, for $q>2$, due to this shortage of   embedded
domains (neglecting the small thermal fluctuations), there is little
difference between geometric domains and hull-enclosed area
measures,  implying that the
parameters of the distributions should have similar values.

\subsubsection{$2< q \leq 4$ and $T_0\to\infty$ (finite correlation length)}

Despite the existence of many similarities between the Potts model
with $q=2, \ 3$ and 4, there are also some fundamental differences.
Besides all having a continuous transition, the equal time
correlation function seems to share the same universal scaling
function and the related correlation length grows with the same power
of time, $t^{1/2}$, see Sect.~\ref{section.Potts}.  Nevertheless,
differently from the  $q=2$ case, that presents a percolating domain
with probability almost one as early as $t=2$ after the
quench~\cite{ArBrCuSi07,SiArBrCu07}, the $q>2$, $T_0\to\infty$ initial
condition is sufficiently far from critical percolation
that the system remains, at least in the time window of our
simulations, distant from the percolation threshold (in spite of the
largest domain steadily, but slowly, increasing with time). 

\begin{figure}[h]
\psfrag{A}{$A$}
\psfrag{A2}{$A^{-2}$}
\psfrag{nh}{$n_h(A,t)$}
\includegraphics[width=8cm]{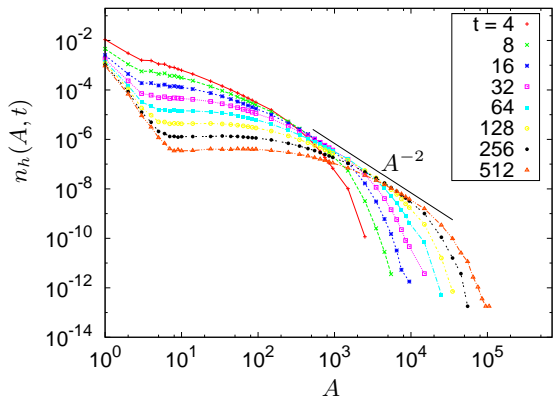}
\caption{(Color online) Hull-enclosed area distribution at
several times (given in the key) after a quench from equilibrium at
$T_0\to\infty$ to $T_f=T_c/2$ in the $q=3$ case. Analogous distributions
are obtained for $q>4$ and $T_0\to\infty$ (not shown). The declivity
of the envelope is $-2$ as a consequence of the scaling obeyed
by the distribution (see text).}
\label{fig.dist_nh_q3_tinf}
\end{figure}

As the
system evolves after the quench, the distribution keeps memory of the
initial state, that corresponds to random percolation with occupation
probability $p=1/q$. And, by not
getting close to a critical point, the distributions do not become
critical and, as a consequence, do not develop a power law tail, as
illustrated in Fig.~\ref{fig.dist_nh_q3_tinf} for $q=3$ and 
$T_0\to \infty$. There is, however, an $A^{-2}$ envelope that is a 
direct consequence of dynamical scaling, present also for other
values of $q$.
The scaling hypothesis requires that the hull-enclosed area
distribution satisfies $n_h(A,t)=t^{-2}n_h(A/t)$. As a consequence,
the envelope straight line shown in the figure should have a $-2$
declivity. To see this, consider two of the curves shown (corresponding
to times $t_1^*$ and $t_2^*$) and define $A^*$ as the location of 
the point that is tangent to the envelope. If we use this value to rescale 
the distributions, we obtain
$R^4(t_1^*)n_1^*=R^4(t_2^*)n_2^*$, where $n_i^*$ is 
$n(A_i^*/R^2(t_i^*))$. That is, $n_2^*/n_1^*=(A_1^*/A_2^*)^2$. Taking
the logarithm of both sides, as in the graph, this gives the $-2$
declivity. The tangent point can thus be an alternative way to obtain
the characteristic length in these systems. 

Very similar distributions (not shown) are obtained for $q>4$ and $T_0=T_c$, another case 
with only finite correlation length in the initial state.

\subsubsection{$2\leq q\leq 4$ and $T_0=T_c$ (infinite correlation length)}

\begin{figure}[h]
\psfrag{A}{$A/t$}
\psfrag{nh}{$t^2n_h(A,t)$}
\psfrag{A2}{$A$}
\psfrag{nh2}{$\!\!\!n_h(A,t)$}
\includegraphics[width=8cm]{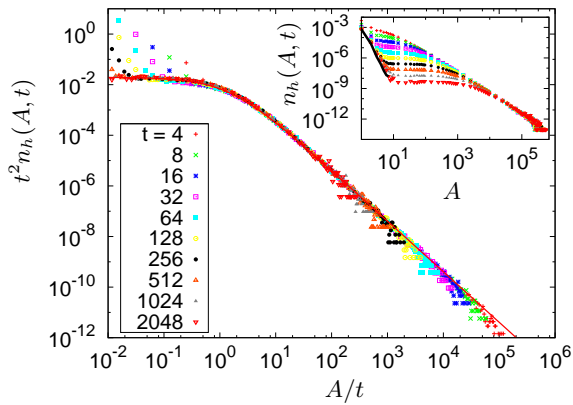}
\caption{(Color online) Collapsed hull-enclosed area distributions at
several times after a quench from equilibrium at
$T_0=T_c$ to $T_f=T_c/2$, for $q=3$.  The line is Eq.~(\ref{eq.dist_nh})
with $\lambda_h^{(3)}\simeq 1.4$. The points at $A/t\ll 1$ that
deviate from the scaling function are due to thermal fluctuations,
depicted as a continuous black line in the inset.}
\label{fig.dist_q3_tc}
\end{figure}

\begin{figure}[h]
\psfrag{dA/dt}{$dA/dt$}
\psfrag{t}{$t$}
\includegraphics[width=8cm]{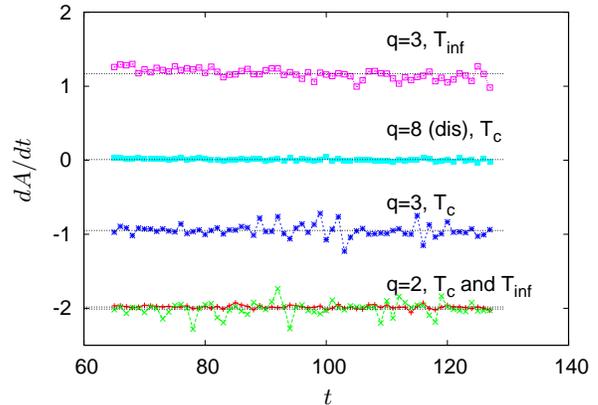}
\caption{(Color online) Average rate of area change as a function of time for
several cases studied in the text. Although the von Neumann law predicts that
each domain has a different rate depending on its number of sides, the
average is constant in time and non-positive for cases with a hull-enclosed and domain area
pdfs with power-law tails.}
\label{fig.vonneumann}
\end{figure}

Figure~\ref{fig.dist_q3_tc} shows the collapsed distribution of
hull-enclosed areas after a quench from $T_0=T_c$ to $T_f=T_c/2$.  We
do not show the related figure for the geometric domains (without any
spanning cluster) since they are almost indistinguishable, another
indication that the parameters appearing in both distributions
differ by very little. The overall behavior is similar to the $q=2$
case, in which the collapse of curves for different times onto a
single universal function demonstrates the existence of a single
length scale that, moreover, follows the Allen-Cahn growth law,
$R(t) \sim t^{1/2}$. Assuming now that the number of sides in 
the von Neumann equation (\ref{eq:Neumann}) 
can be replaced by a constant mean, $n \to \langle n\rangle$, 
and using Eq.~(\ref{eq.tct0}) and the results in 
Refs.~\cite{ArBrCuSi07,SiArBrCu07},
the hull-enclosed area distribution for $q\leq 4$, within this
mean-field-like approximation becomes
\begin{equation}
n_h(A,t) = \frac{(q-1)c_h^{(q)}}{\left(A + \lambda_h^{(q)} t\right)^2},
\label{eq.dist_nh}
\end{equation}
that fits very accurately the data using $\lambda_h^{(3)}\simeq 1.4$. 
The deviations present at small values of $A/t$ are due to
thermal fluctuations that are visible in the inset (see \cite{SiArBrCu07}
for details). 
The above mean-field-like approximation can be
directly tested by measuring the average change in area, $dA/dt$, 
that is, the number of spins included or excluded in those domains
that survived during a given time interval. Figure~\ref{fig.vonneumann}
shows the results for several cases. Although the von Neumann law predicts that
each domain has a different rate, either positive or negative, depending on its 
number of sides, the average, effective $\lambda$ is constant, as is the case 
for the Ising model $q=2$. Similarly, only those that have 
$\lambda_{\scriptscriptstyle\rm eff}\leq 0$  present a power 
law distribution. Interestingly,
the case with $q=8$ with ferromagnetic disorder, to be discussed in
section~\ref{section.disorder}, seems to be marginal, 
$\lambda_{\scriptscriptstyle\rm eff}\simeq 0$. Thus, there seems to be
a net difference between cases that give a power law pdf the ones that 
do not. The detailed implications, that probably involve
the knowledge of perimeter and number of sides pdfs are beyond the
scope of this paper and we postpone their study to a future work.

\begin{figure}[h]
\psfrag{A}{$A$}
\psfrag{A2}{$A^{-2}$}
\psfrag{nh}{$n_h(A,t)$}
\includegraphics[width=8cm]{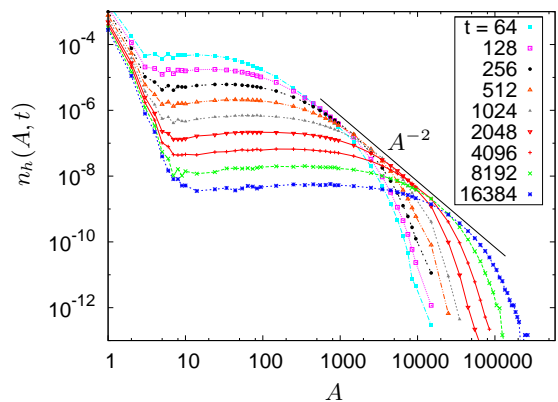}
\caption{(Color online)
Hull-enclosed area distribution after a quench from  
$T_0=T_c$ to $T_f=T_c/2$ in the $q=8$ model at different times
given in the key. }
\label{fig.dist_nh_q8_tc}
\end{figure}

\subsubsection{$q > 4$  (finite correlation length)}
\label{sec.tcq8}

For $q>4$, no initial state is critical and the initial
distributions are not power law. After the quench, the system
keeps some memory of this initial state and the scaling state has a
more complex universal function.
Figure~\ref{fig.dist_nh_q8_tc} exemplifies this
behavior for $q=8$ and $T_0=T_c$. Again, since the quench is to
$T_c/2$ the small $A$ region, almost time independent, is due to the
large number of small clusters formed by thermal fluctuations. After
this initial decaying region, the distribution increases with $A$ (the
width of such region increases with time), to then decrease
exponentially.  As for a quench from $T_0\to
\infty$ and $q=3$, an envelope power law ($A^{-2}$)
forms when several distributions for different values of $t$ are
considered, and this is a direct consequence of dynamical scaling.
The precise analytic form of the scaling function is not known and
has been a matter of debate for several decades not only for the Potts
model but for related models of interest for the grain growth
community (see \cite{FrUd94,Mullins98,RiLu01,Flyvbjerg93a} and references therein),
this issue being still unsettled. 

\begin{figure}[h]
\psfrag{AR2}{$A/R^2(t)$}
\psfrag{R4n}{\hspace{-5mm}$AR^2(t) n_h(A,t)$}
\psfrag{nh2}{\hspace{-5mm}$R^4(t) n_h(A,t)$}
\includegraphics[width=8cm]{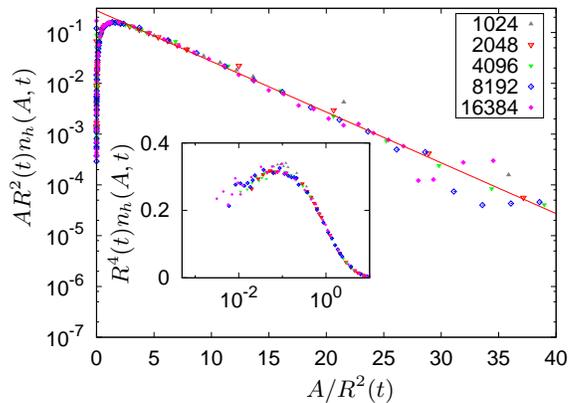}
\caption{(Color online) Dynamical scaling of the distributions shown
in Fig.~\ref{fig.dist_nh_q8_tc} for $q=8$ and $T_0=T_c$.
A good collapse is obtained, for $A>R^2(t)$, with
$n_h(A,t)=R^{-4}(t)f(A/R^2)$ (the data for $A<10$, corresponding to
thermal fluctuations, have been removed from the inset).  The main
panel shows the exponential tail, $f(x)\sim x^{-1}\exp(-ax)$, where
the fitting parameter is $a\simeq 0.23$. Data for $T_0\to\infty$ collapse
onto the same universal function (not shown). The inset also shows 
the small area region, where the collapse is not as good, probably due
to thermal interface roughening.}
\label{fig.dist_nh_q8_tc_colr2}
\end{figure}

The tail keeps memory of the initial
condition and is exponential for $q>4$. This fact is better appreciated in 
Fig.~\ref{fig.dist_nh_q8_tc_colr2} where we use a scaling description of the data in 
Fig.~\ref{fig.dist_nh_q8_tc} and a linear-log scale. The universal scaling function
shown in the inset, for $A>R^2(t)$, is the same for both $T_0=T_c$
and $T_0\to\infty$ (not shown): $f(x)\sim x^{-1}\exp(-ax)$, with a fitting parameter
$a\simeq 0.23$ in both cases. This is a consequence of the finiteness of the
correlation length at $t=0$ and the fact that the tail of the
distribution samples large areas. Notice also that the collapse is worse
for small values of $A/R^2(t)$, probably due to the temperature roughening
of the domain walls.

\section{Weak quenched disorder}
\label{section.disorder}

It is also possible to study the coarsening behavior in the
presence of random ferromagnetic (weak) disorder
obtained, for example, by choosing the bonds from a probability 
distribution function, $P(J_{ij})$, 
with semi-definite positive support, $J_{ij}\geq 0$. Weak 
disorder weakens the phase transition that, in the $2d$ random bond 
Potts model (RBPM)
becomes continuous for all $q$~\cite{HuBe89,AiWe89,ChFeLa92,Ludwig90,BeCh04}.
For the bimodal distribution 
\begin{equation}
P(J_{ij}) = \frac{1}{2}\delta(J_{ij}-J_1) + \frac{1}{2}\delta(J_{ij}-J_2),
\end{equation}
that we shall treat here, the transition occurs at~\cite{KiDo81}
\begin{equation}
\left(e^{\beta_c J_1}-1\right)\left(e^{\beta_c J_2}-1\right)=q.
\label{eq.disordertc}
\end{equation}
  We use  $J_1=1$ and $J_2=1/2$, that
is, $J_1/J_2=2$, for which  $T_c\simeq 1.443$ for $q=3$ and
$T_c\simeq 1.087$ for $q=8$.
The additional  contribution to interface pinning due to quenched disorder
may be useful to understand phenomena such as the so-called
Zener pinning~\cite{SrGr85,HaOl90,KrSt92}.

\begin{figure}[h]
\psfrag{r}{$r/R(t)$}
\psfrag{C}{$C(r,t)$}
\psfrag{a}{$r^{-\eta}$}
\psfrag{A2}{$A$}
\psfrag{nh2}{$\hspace{-4mm}n_h(A,0)$}
\includegraphics[width=9cm]{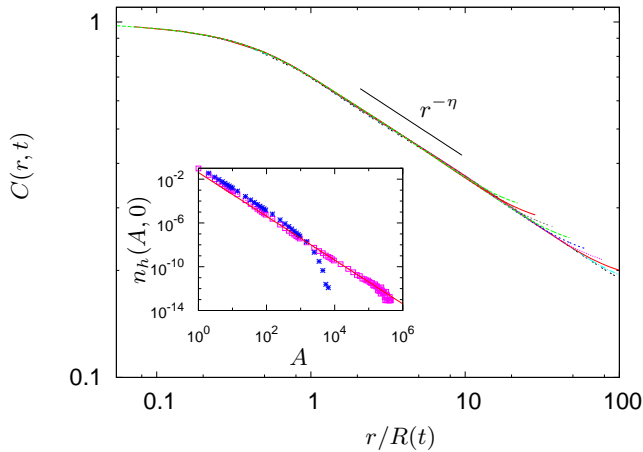}
\caption{(Color online) 
Rescaled space-time correlation function at several times
($t=2^{2},\ldots,2^{10}$ MCs) after a quench from $T_0=T_c$ to 
$T_f=T_c/2$ for $q=8$ and bimodal disorder in log-log scale
(compare with the pure $q=8$ case in Fig.~\ref{fig.crtc_q8}).
The tail keeps memory of the $r^{-\eta}$ decay of the initial
equilibrium state. We found that $\eta\simeq 0.28$ fits well the
data. A similar figure is obtained for $q=3$, with an
exponent close to $0.27$. Inset: equilibrium distribution of
hull-enclosed areas at $T_c$ with weak disorder (red) -- a critical 
case with long-range correlations -- and
without (blue) -- a case with short-range correlations.  
The initial distribution was obtained after 3000 SW
steps and is indistinguishable, within the errors, from the distribution of 
the pure model with $q=3$ (line).}
\label{fig.crtc_q8dis_col}
\end{figure}

The growth law for the random $q=2$ case, whether logarithmic or a
power-law with a $T$ and disorder-strength dependent exponent, has
been a subject of debate~\cite{PaPuRi04,PaPuRi05,HePl06,PaScRi07} and
arguments for a crossover between the pure growth law $t^{1/2}$ to
logarithmic growth at an equilibrium length-scale -- that can be
easily confused with the existence of $T$ and disorder dependent exponent 
in a power-law growth -- were
recently given in~\cite{Igbukocu09}. Whether the asymptotic
logarithmic growth also applies in the case $q>2$ has not been tested
numerically.  Regardless of the growth law, dynamical
scaling is observed in the correlation and area distribution 
functions for $q=2$~\cite{SiArBrCu08} as well as for $q>2$.
The rescaled space-time correlations for the disordered $q=3$ and $q=8$ 
after a quench from $T_0\to\infty$ are indistinguishable from 
the pure case, see Fig.~\ref{fig.crtinf}. Notice
that the initial condition, being random, is not affected by the
presence of disorder, differently from the $T_0=T_c$ case, 
where the disorder changes the nature (correlations) of the initial states, 
leading to a failure of the super-universality hypothesis. In this case,
 long range
correlations are present and the decay is power-law, $C(r)\sim r^{-\eta}$,
as shown in Fig.~\ref{fig.crtc_q8dis_col}, in analogy with the pure $2\leq q\leq 4$ 
cases. However, as it was pointed out in Refs.~\cite{Picco98,OlYo99}, 
 for the 
bimodal distribution and the values of $J_i$ chosen here, this may 
suffer from crossover effects, not yet being in the asymptotic
scaling regime. For  $q=3$ and 8, the fitted exponent is $\eta\simeq 0.27$
and 0.28, respectively, results that are compatible with 
Refs.~\cite{JaCa98,Picco98}.

\begin{figure}[h]
\psfrag{AR2}{$A/R^2(t)$}
\psfrag{R4n}{\hspace{-5mm}$R^4(t)n_h(A,t)$}
\psfrag{NR4}{\hspace{-5mm}$R^4(t)n_h(A,t)$}
\includegraphics[width=8cm]{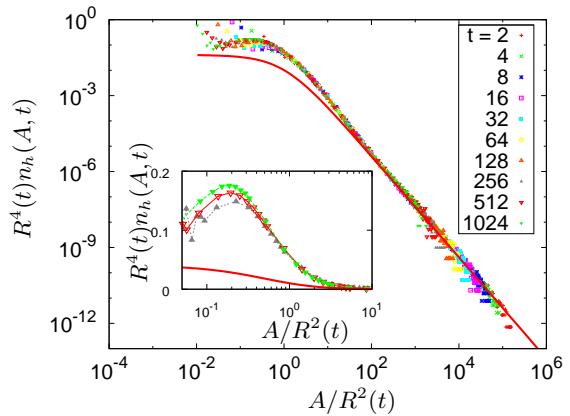}
\caption{(Color online)
Collapsed hull-enclosed area distribution for $q=8$ after a
quench from $T_0=T_c$ to $T_f=T_c/2$ in the RBPM  (half the bonds are changed from 1 to
0.5). Data taken at several times are shown with different symbols.  
The thick line shown is $2c_h^{(3)}/(1+x)^2$ (the pure $q=3$ case).
The tail keeps memory of the initial state and
is well described by a power law (solid red line) with exponent $-2$.
 Inset: the small $A$ region does not scale and its
behavior is reminiscent of what is seen in the pure case at small areas
(but narrower), see the inset in Fig.~\ref{fig.dist_nh_q8_tc_colr2}.}
\label{fig.dist_q8dis_tc}
\end{figure}

Figure~\ref{fig.dist_q8dis_tc} shows the rescaled hull-enclosed
area distribution function for the disordered $q=8$ model quenched from 
equilibrium at $T_0=T_c$. 
The transition is continuous for all $q$ and the equilibrium 
area distribution at $T_c$ is described, for large $A$ and
within the simulation error, by the same power law as the 
pure model with $q=3$, Eq.~(\ref{eq.tct0}), see  the 
inset of Fig.~\ref{fig.crtc_q8dis_col}. There are, however, 
small deviations for not so large values
of $A$ that, after the quench, become more apparent, as can be
observed in the main panel of Fig.~\ref{fig.dist_q8dis_tc}, 
and in the inset where a zoom on this region is 
shown. This may be an effect of
pinning by disorder that keeps an excess of small domains, slowing
down their evolution, and not letting dynamic scaling establish at
these scales. The weak increase of the pdf at small $A$s, where the
type of dynamics is important, resembles also the
behavior of the pure case, see the inset in 
Fig.~\ref{fig.dist_nh_q8_tc_colr2}.

\begin{figure}[h]
\psfrag{AR2}{$A/R^2(t)$}
\psfrag{R4n}{\hspace{-5mm}$R^4(t)n_h(A,t)$}
\psfrag{A2}{$A$}
\psfrag{nh2}{$\hspace{-4mm}n_h(A,0)$}
\includegraphics[width=8cm]{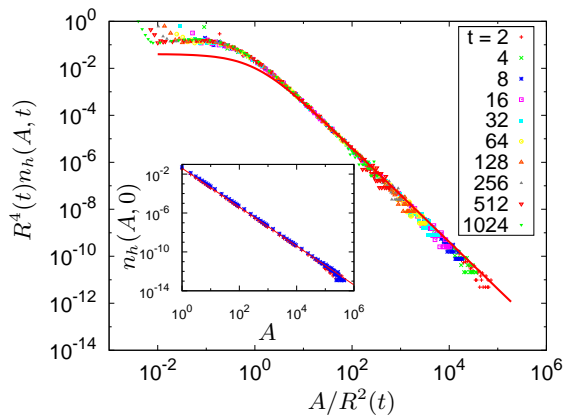}
\caption{(Color online)
Collapsed hull-enclosed area distribution for $q=3$ after a
quench from $T_0=T_c$ to $T_f=T_c/2$ in the RBPM  (half the bonds are changed from 1 to
0.5). Differently from the $q=8$ case shown in Fig.~\ref{fig.dist_q8dis_tc},
here both the pure and disordered initial state are critical (inset),
and the distributions at $t=0$ are indistinguishable. Nonetheless, the
subsequent evolution of these cases are not equivalent.
For roughly $A<R^2(t)$ the scaling function for the disordered
case no longer follows the pure case function, $2c_h^{(3)}/(1+x)^2$. 
Interestingly, this equation describes the tail for both $q=3$ and
$8$ (previous figure).}
\label{fig.dist_q3dis_tc}
\end{figure}

Interestingly enough, the long-range correlations in the
$T_c$ initial condition, that in presence of quenched 
disorder exist for all $q$, determine the large area behavior of the 
dynamic pdf, more precisely, they dictate their power-law decay.
Thus, for $q>4$ pure and disordered systems behave 
qualitatively differently when the initial conditions are in equilibrium at
$T_0=T_c$. An interesting characteristic of both figs.~\ref{fig.dist_q8dis_tc} 
for $q=8$ and \ref{fig.dist_q3dis_tc} for $q=3$ is that the tail is well 
described by the pure distribution with $q=3$.

For $q\leq 4$, the initial state at $T_c$ has long-range correlations with and
without disorder. We study the effect of disorder on the hull-enclosed area 
distribution in Fig.~\ref{fig.dist_q3dis_tc}. The tail is a power law compatible with 
the $-2$ expectation. The small area behavior is interesting: the data
deviates from the critical behavior as can be seen in the
inset of Fig.~\ref{fig.dist_q8dis_tc}, showing a non-monotonic
excess of small domains that resembles the behavior of the pure case. Somehow, disorder,
even if local, has a strong effect at large scales while affecting
less the properties at small ones.

\section{Conclusions}
\label{section.conclusions}

We presented a systematic study of some geometric properties of
the Potts model during the coarsening dynamics after a sudden
quench in temperature. Although the theory for the Ising
model ($q=2$) cannot be easily extended to $q>2$, our numerical results
are a first step in this direction. 

Our analysis demonstrates the fundamental role played by the initial conditions,
more precisely, whether they have an infinite correlation length or not. 

The distribution of hull-enclosed areas 
in pure $2d$ Potts models with $2 \leq  q \leq 4$ and dirty cases with all $q$ 
that evolved from initial conditions in equilibrium at $T_c$, that is to 
say cases with an infinite correlation length, are large, with power-law
tails that are compatible with the exponent $-2$ within our numerical 
accuracy. The small area behavior is richer. In 
cases with $2\leq q\leq 4$, again
within our numerical accuracy, the hull-enclosed area is well captured by a simple extension 
of the distribution in Eq.~(\ref{eq.dist_nh_d}):
\begin{equation}
n_h(A,t) \simeq \frac{(q-1)c_h^{(q,d)}}{\left[A + R^2(t)\right]^2}
,
\qquad\qquad 2\leq q \leq 4
,
\label{eq.dist_nh_d}
\end{equation}
 that upgrades the prefactor $c_h$ to depend on 
$q$ and disorder. In disordered cases with $q>4$ this form does not 
describe the small $A$ dependence. Indeed, the scaled pdf has
a non monotonic behavior both with and without disorder, with
dynamic scaling failing in the explored times, possibly due to strong 
pinning effects.

Very different are the distributions of dynamic hull-enclosed areas
evolved from initial states with finite correlation lengths, 
as those obtained in equilibrium at  $T_0\to\infty$.
The scaling functions of the dynamic distributions 
are reminiscent of the disordered state at the initial temperature
with an exponential tail. Differently from the Ising case, in which 
the percolation critical point gave a long-tail to the low-temperature 
distribution, in cases with $q>2$ this does not occur.

There are further geometrical properties that were not explored here
and  deserve attention. For example, the distribution of perimeter
lengths, number of sides, and the correlation between area and
number of sides are of interest in cellular systems. In particular,
it would be interesting to check whether the Aboav-Weaire and Lewis 
law~\cite{Schliecker02} are valid in the $2d$ Potts model with and 
without random bonds.

Finally, the Potts model is realized in a few experimental situations.
Our results should be a guideline to search for 
similar  distributions, 
in analogy to what has been done in the $q=2$
case, by using a liquid crystal sample~\cite{Si08}.

\begin{acknowledgments}
We acknowledge fruitful conversations with S. A. Cannas, M. J. de Oliveira, 
M. Picco, 
D. A. Stariolo and R. Ziff. We specially thank A. J. Bray for
his early collaboration in the project. Work partially supported by
a CAPES/Cofecub grant 448/04.
JJA and MPOL are partially supported by the Brazilian agency CNPq.  
\end{acknowledgments}

\bibliographystyle{apsrev}

\begin{thebibliography}{69}
\expandafter\ifx\csname natexlab\endcsname\relax\def\natexlab#1{#1}\fi
\expandafter\ifx\csname bibnamefont\endcsname\relax
  \def\bibnamefont#1{#1}\fi
\expandafter\ifx\csname bibfnamefont\endcsname\relax
  \def\bibfnamefont#1{#1}\fi
\expandafter\ifx\csname citenamefont\endcsname\relax
  \def\citenamefont#1{#1}\fi
\expandafter\ifx\csname url\endcsname\relax
  \def\url#1{\texttt{#1}}\fi
\expandafter\ifx\csname urlprefix\endcsname\relax\def\urlprefix{URL }\fi
\providecommand{\bibinfo}[2]{#2}
\providecommand{\eprint}[2][]{\url{#2}}

\bibitem[{\citenamefont{Glazier et~al.}(1990)\citenamefont{Glazier, Anderson,
  and Grest}}]{GlAnGr90}
\bibinfo{author}{\bibfnamefont{J.~A.} \bibnamefont{Glazier}},
  \bibinfo{author}{\bibfnamefont{M.~P.} \bibnamefont{Anderson}},
  \bibnamefont{and} \bibinfo{author}{\bibfnamefont{G.~S.} \bibnamefont{Grest}},
  \bibinfo{journal}{Phil. Mag. B} \textbf{\bibinfo{volume}{62}},
  \bibinfo{pages}{615} (\bibinfo{year}{1990}).

\bibitem[{\citenamefont{Mombach et~al.}(1993)\citenamefont{Mombach, de~Almeida,
  and Iglesias}}]{MoAlIg93}
\bibinfo{author}{\bibfnamefont{J.~C.~M.} \bibnamefont{Mombach}},
  \bibinfo{author}{\bibfnamefont{R.~M.~C.} \bibnamefont{de~Almeida}},
  \bibnamefont{and} \bibinfo{author}{\bibfnamefont{J.~R.}
  \bibnamefont{Iglesias}}, \bibinfo{journal}{Phys. Rev. E}
  \textbf{\bibinfo{volume}{48}}, \bibinfo{pages}{598} (\bibinfo{year}{1993}).

\bibitem[{\citenamefont{Prozorov et~al.}(2008)\citenamefont{Prozorov, Fidler,
  Hoberg, and Canfield}}]{PrFiHoCa08}
\bibinfo{author}{\bibfnamefont{R.}~\bibnamefont{Prozorov}},
  \bibinfo{author}{\bibfnamefont{A.~F.} \bibnamefont{Fidler}},
  \bibinfo{author}{\bibfnamefont{J.~R.} \bibnamefont{Hoberg}},
  \bibnamefont{and} \bibinfo{author}{\bibfnamefont{P.~C.}
  \bibnamefont{Canfield}}, \bibinfo{journal}{Nature Physics}
  \textbf{\bibinfo{volume}{4}}, \bibinfo{pages}{327} (\bibinfo{year}{2008}).

\bibitem[{\citenamefont{Babcock et~al.}(1990)\citenamefont{Babcock, Seshadri,
  and Westervelt}}]{BaSeWe90}
\bibinfo{author}{\bibfnamefont{K.~L.} \bibnamefont{Babcock}},
  \bibinfo{author}{\bibfnamefont{R.}~\bibnamefont{Seshadri}}, \bibnamefont{and}
  \bibinfo{author}{\bibfnamefont{R.~M.} \bibnamefont{Westervelt}},
  \bibinfo{journal}{Phys. Rev. A} \textbf{\bibinfo{volume}{41}},
  \bibinfo{pages}{1952} (\bibinfo{year}{1990}).

\bibitem[{\citenamefont{Jagla}(2004)}]{Jagla04}
\bibinfo{author}{\bibfnamefont{E.~A.} \bibnamefont{Jagla}},
  \bibinfo{journal}{Phys. Rev. E} \textbf{\bibinfo{volume}{70}},
  \bibinfo{pages}{046204} (\bibinfo{year}{2004}).

\bibitem[{\citenamefont{Mullins}(1956)}]{Mullins56}
\bibinfo{author}{\bibfnamefont{W.~W.} \bibnamefont{Mullins}},
  \bibinfo{journal}{J. Appl. Phys.} \textbf{\bibinfo{volume}{27}},
  \bibinfo{pages}{900} (\bibinfo{year}{1956}).

\bibitem[{\citenamefont{Arenzon et~al.}(2007)\citenamefont{Arenzon, Bray,
  Cugliandolo, and Sicilia}}]{ArBrCuSi07}
\bibinfo{author}{\bibfnamefont{J.~J.} \bibnamefont{Arenzon}},
  \bibinfo{author}{\bibfnamefont{A.~J.} \bibnamefont{Bray}},
  \bibinfo{author}{\bibfnamefont{L.~F.} \bibnamefont{Cugliandolo}},
  \bibnamefont{and} \bibinfo{author}{\bibfnamefont{A.}~\bibnamefont{Sicilia}},
  \bibinfo{journal}{Phys. Rev. Lett.} \textbf{\bibinfo{volume}{98}},
  \bibinfo{pages}{145701} (\bibinfo{year}{2007}).

\bibitem[{\citenamefont{von Neumann}(1952)}]{Neumann52}
\bibinfo{author}{\bibfnamefont{J.}~\bibnamefont{von Neumann}}, in
  \emph{\bibinfo{booktitle}{Metal Interfaces}}, edited by
  \bibinfo{editor}{\bibfnamefont{C.}~\bibnamefont{Herring}}
  (\bibinfo{organization}{American Society for Metals}, \bibinfo{year}{1952}),
  pp. \bibinfo{pages}{108--110}.

\bibitem[{\citenamefont{Glazier and Stavans}(1989)}]{GlSt89}
\bibinfo{author}{\bibfnamefont{J.~A.} \bibnamefont{Glazier}} \bibnamefont{and}
  \bibinfo{author}{\bibfnamefont{J.}~\bibnamefont{Stavans}},
  \bibinfo{journal}{Phys. Rev. A} \textbf{\bibinfo{volume}{40}},
  \bibinfo{pages}{7398} (\bibinfo{year}{1989}).

\bibitem[{\citenamefont{MacPherson and Srolovitz}(2007)}]{MaSr07}
\bibinfo{author}{\bibfnamefont{R.~D.} \bibnamefont{MacPherson}}
  \bibnamefont{and} \bibinfo{author}{\bibfnamefont{D.~J.}
  \bibnamefont{Srolovitz}}, \bibinfo{journal}{Nature}
  \textbf{\bibinfo{volume}{446}}, \bibinfo{pages}{1053} (\bibinfo{year}{2007}).

\bibitem[{\citenamefont{Grest et~al.}(1988)\citenamefont{Grest, Anderson, and
  Srolovitz}}]{GrAnSr88}
\bibinfo{author}{\bibfnamefont{G.~S.} \bibnamefont{Grest}},
  \bibinfo{author}{\bibfnamefont{M.~P.} \bibnamefont{Anderson}},
  \bibnamefont{and} \bibinfo{author}{\bibfnamefont{D.~J.}
  \bibnamefont{Srolovitz}}, \bibinfo{journal}{Phys. Rev. B}
  \textbf{\bibinfo{volume}{38}}, \bibinfo{pages}{4752} (\bibinfo{year}{1988}).

\bibitem[{\citenamefont{Lau et~al.}(1988)\citenamefont{Lau, Dasgupta, and
  Valls}}]{LaDaVa88}
\bibinfo{author}{\bibfnamefont{M.}~\bibnamefont{Lau}},
  \bibinfo{author}{\bibfnamefont{C.}~\bibnamefont{Dasgupta}}, \bibnamefont{and}
  \bibinfo{author}{\bibfnamefont{O.~T.} \bibnamefont{Valls}},
  \bibinfo{journal}{Phys. Rev. B} \textbf{\bibinfo{volume}{38}},
  \bibinfo{pages}{9024} (\bibinfo{year}{1988}).

\bibitem[{\citenamefont{Lifshitz}(1962)}]{Lifshitz62}
\bibinfo{author}{\bibfnamefont{I.~M.} \bibnamefont{Lifshitz}},
  \bibinfo{journal}{Sov. Phys. JETP} \textbf{\bibinfo{volume}{15}},
  \bibinfo{pages}{939} (\bibinfo{year}{1962}).

\bibitem[{\citenamefont{Safran}(1981)}]{Safran81}
\bibinfo{author}{\bibfnamefont{S.~A.} \bibnamefont{Safran}},
  \bibinfo{journal}{Phys. Rev. Lett.} \textbf{\bibinfo{volume}{46}},
  \bibinfo{pages}{1581} (\bibinfo{year}{1981}).

\bibitem[{\citenamefont{Wu}(1982)}]{Wu82}
\bibinfo{author}{\bibfnamefont{F.~Y.} \bibnamefont{Wu}}, \bibinfo{journal}{Rev.
  Mod. Phys.} \textbf{\bibinfo{volume}{54}}, \bibinfo{pages}{235}
  (\bibinfo{year}{1982}).

\bibitem[{\citenamefont{de~Oliveira et~al.}(2004)\citenamefont{de~Oliveira,
  Petri, and Tomé}}]{OlPeTo04}
\bibinfo{author}{\bibfnamefont{M.~J.} \bibnamefont{de~Oliveira}},
  \bibinfo{author}{\bibfnamefont{A.}~\bibnamefont{Petri}}, \bibnamefont{and}
  \bibinfo{author}{\bibfnamefont{T.}~\bibnamefont{Tomé}},
  \bibinfo{journal}{Europhys. Lett.} \textbf{\bibinfo{volume}{65}},
  \bibinfo{pages}{20} (\bibinfo{year}{2004}).

\bibitem[{\citenamefont{Ferrero and Cannas}(2007)}]{FeCa07}
\bibinfo{author}{\bibfnamefont{E.~E.} \bibnamefont{Ferrero}} \bibnamefont{and}
  \bibinfo{author}{\bibfnamefont{S.~A.} \bibnamefont{Cannas}},
  \bibinfo{journal}{Phys. Rev. E} \textbf{\bibinfo{volume}{76}},
  \bibinfo{pages}{031108} (\bibinfo{year}{2007}).

\bibitem[{\citenamefont{de~Berganza et~al.}(2007)\citenamefont{de~Berganza,
  Ferrero, Cannas, Loreto, and Petri}}]{BeFeCaLoPe07}
\bibinfo{author}{\bibfnamefont{M.~I.} \bibnamefont{de~Berganza}},
  \bibinfo{author}{\bibfnamefont{E.~E.} \bibnamefont{Ferrero}},
  \bibinfo{author}{\bibfnamefont{S.~A.} \bibnamefont{Cannas}},
  \bibinfo{author}{\bibfnamefont{V.}~\bibnamefont{Loreto}}, \bibnamefont{and}
  \bibinfo{author}{\bibfnamefont{A.}~\bibnamefont{Petri}},
  \bibinfo{journal}{Eur. Phys. J. Special Topics}
  \textbf{\bibinfo{volume}{143}}, \bibinfo{pages}{273} (\bibinfo{year}{2007}).

\bibitem[{\citenamefont{Sicilia et~al.}(2007)\citenamefont{Sicilia, Arenzon,
  Bray, and Cugliandolo}}]{SiArBrCu07}
\bibinfo{author}{\bibfnamefont{A.}~\bibnamefont{Sicilia}},
  \bibinfo{author}{\bibfnamefont{J.~J.} \bibnamefont{Arenzon}},
  \bibinfo{author}{\bibfnamefont{A.~J.} \bibnamefont{Bray}}, \bibnamefont{and}
  \bibinfo{author}{\bibfnamefont{L.~F.} \bibnamefont{Cugliandolo}},
  \bibinfo{journal}{Phys. Rev. E} \textbf{\bibinfo{volume}{76}},
  \bibinfo{pages}{061116} (\bibinfo{year}{2007}).

\bibitem[{\citenamefont{Sicilia et~al.}(2009)\citenamefont{Sicilia, Sarrazin,
  Arenzon, Bray, and Cugliandolo}}]{SiSaArNrCu09}
\bibinfo{author}{\bibfnamefont{A.}~\bibnamefont{Sicilia}},
  \bibinfo{author}{\bibfnamefont{Y.}~\bibnamefont{Sarrazin}},
  \bibinfo{author}{\bibfnamefont{J.~J.} \bibnamefont{Arenzon}},
  \bibinfo{author}{\bibfnamefont{A.~J.} \bibnamefont{Bray}}, \bibnamefont{and}
  \bibinfo{author}{\bibfnamefont{L.~F.} \bibnamefont{Cugliandolo}},
  \bibinfo{journal}{Phys. Rev. E} \textbf{\bibinfo{volume}{80}},
  \bibinfo{pages}{031121} (\bibinfo{year}{2009}).

\bibitem[{\citenamefont{Cardy and Ziff}(2003)}]{CaZi03}
\bibinfo{author}{\bibfnamefont{J.}~\bibnamefont{Cardy}} \bibnamefont{and}
  \bibinfo{author}{\bibfnamefont{R.~M.} \bibnamefont{Ziff}},
  \bibinfo{journal}{J. Stat. Phys.} \textbf{\bibinfo{volume}{110}},
  \bibinfo{pages}{1} (\bibinfo{year}{2003}).

\bibitem[{\citenamefont{Barros et~al.}(2009)\citenamefont{Barros, Krapivsky,
  and Redner}}]{BaKrRe09}
\bibinfo{author}{\bibfnamefont{K.}~\bibnamefont{Barros}},
  \bibinfo{author}{\bibfnamefont{P.}~\bibnamefont{Krapivsky}},
  \bibnamefont{and} \bibinfo{author}{\bibfnamefont{S.}~\bibnamefont{Redner}},
  \bibinfo{journal}{Phys. Rev. E} \textbf{\bibinfo{volume}{80}},
  \bibinfo{pages}{040101(R)} (\bibinfo{year}{2009}).

\bibitem[{\citenamefont{Newman and Barkema}(1999)}]{NeBa99}
\bibinfo{author}{\bibfnamefont{M.}~\bibnamefont{Newman}} \bibnamefont{and}
  \bibinfo{author}{\bibfnamefont{G.}~\bibnamefont{Barkema}},
  \emph{\bibinfo{title}{Monte Carlo methods in statistical physics}}
  (\bibinfo{publisher}{Oxford University Press}, \bibinfo{address}{New York,
  USA}, \bibinfo{year}{1999}).

\bibitem[{\citenamefont{Loscar et~al.}(2009)\citenamefont{Loscar, Ferrero,
  Grigera, and Cannas}}]{LoFrGrCa09}
\bibinfo{author}{\bibfnamefont{E.~S.} \bibnamefont{Loscar}},
  \bibinfo{author}{\bibfnamefont{E.~E.} \bibnamefont{Ferrero}},
  \bibinfo{author}{\bibfnamefont{T.~S.} \bibnamefont{Grigera}},
  \bibnamefont{and} \bibinfo{author}{\bibfnamefont{S.~A.}
  \bibnamefont{Cannas}}, \bibinfo{journal}{J. Chem. Phys.}
  \textbf{\bibinfo{volume}{131}}, \bibinfo{pages}{024120}
  (\bibinfo{year}{2009}).

\bibitem[{\citenamefont{Kaski et~al.}(1985)\citenamefont{Kaski, Nieminen, and
  Gunton}}]{KaNiGu85}
\bibinfo{author}{\bibfnamefont{K.}~\bibnamefont{Kaski}},
  \bibinfo{author}{\bibfnamefont{J.}~\bibnamefont{Nieminen}}, \bibnamefont{and}
  \bibinfo{author}{\bibfnamefont{J.~D.} \bibnamefont{Gunton}},
  \bibinfo{journal}{Phys. Rev. B} \textbf{\bibinfo{volume}{31}},
  \bibinfo{pages}{2998} (\bibinfo{year}{1985}).

\bibitem[{\citenamefont{Kumar et~al.}(1987)\citenamefont{Kumar, Gunton, and
  Kaski}}]{KuGuKa87}
\bibinfo{author}{\bibfnamefont{S.}~\bibnamefont{Kumar}},
  \bibinfo{author}{\bibfnamefont{J.~D.} \bibnamefont{Gunton}},
  \bibnamefont{and} \bibinfo{author}{\bibfnamefont{K.~K.} \bibnamefont{Kaski}},
  \bibinfo{journal}{Phys. Rev. B} \textbf{\bibinfo{volume}{35}},
  \bibinfo{pages}{8517} (\bibinfo{year}{1987}).

\bibitem[{\citenamefont{Liu and Mazenko}(1993)}]{LiMa93}
\bibinfo{author}{\bibfnamefont{F.}~\bibnamefont{Liu}} \bibnamefont{and}
  \bibinfo{author}{\bibfnamefont{G.~F.} \bibnamefont{Mazenko}},
  \bibinfo{journal}{Phys. Rev. B} \textbf{\bibinfo{volume}{47}},
  \bibinfo{pages}{2866} (\bibinfo{year}{1993}).

\bibitem[{\citenamefont{Rapapa and Maliehe}(2005)}]{RaMa05}
\bibinfo{author}{\bibfnamefont{N.~P.} \bibnamefont{Rapapa}} \bibnamefont{and}
  \bibinfo{author}{\bibfnamefont{N.~B.} \bibnamefont{Maliehe}},
  \bibinfo{journal}{Eur. Phys. J. B} \textbf{\bibinfo{volume}{48}},
  \bibinfo{pages}{219} (\bibinfo{year}{2005}).

\bibitem[{\citenamefont{Humayun and Bray}(1992)}]{HuBr92}
\bibinfo{author}{\bibfnamefont{K.}~\bibnamefont{Humayun}} \bibnamefont{and}
  \bibinfo{author}{\bibfnamefont{A.~J.} \bibnamefont{Bray}},
  \bibinfo{journal}{Phys. Rev. B} \textbf{\bibinfo{volume}{46}},
  \bibinfo{pages}{10594} (\bibinfo{year}{1992}).

\bibitem[{\citenamefont{Bray}(1994)}]{Bray94}
\bibinfo{author}{\bibfnamefont{A.~J.} \bibnamefont{Bray}},
  \bibinfo{journal}{Adv. Phys.} \textbf{\bibinfo{volume}{43}},
  \bibinfo{pages}{357} (\bibinfo{year}{1994}).

\bibitem[{\citenamefont{Humayun and Bray}(1991)}]{HuBr91}
\bibinfo{author}{\bibfnamefont{K.}~\bibnamefont{Humayun}} \bibnamefont{and}
  \bibinfo{author}{\bibfnamefont{A.~J.} \bibnamefont{Bray}},
  \bibinfo{journal}{J. Phys. A: Math. Gen.} \textbf{\bibinfo{volume}{24}},
  \bibinfo{pages}{1915} (\bibinfo{year}{1991}).

\bibitem[{\citenamefont{Buffenoir and Wallon}(1993)}]{BuWa93}
\bibinfo{author}{\bibfnamefont{E.}~\bibnamefont{Buffenoir}} \bibnamefont{and}
  \bibinfo{author}{\bibfnamefont{S.}~\bibnamefont{Wallon}},
  \bibinfo{journal}{J. Phys. A} \textbf{\bibinfo{volume}{26}},
  \bibinfo{pages}{3045} (\bibinfo{year}{1993}).

\bibitem[{\citenamefont{Janke and Kappler}(1995)}]{JaKa95}
\bibinfo{author}{\bibfnamefont{W.}~\bibnamefont{Janke}} \bibnamefont{and}
  \bibinfo{author}{\bibfnamefont{S.}~\bibnamefont{Kappler}},
  \bibinfo{journal}{Phys. Lett. A} \textbf{\bibinfo{volume}{197}},
  \bibinfo{pages}{227} (\bibinfo{year}{1995}).

\bibitem[{\citenamefont{Jacobsen and Picco}(2000)}]{JaPi00}
\bibinfo{author}{\bibfnamefont{J.~L.} \bibnamefont{Jacobsen}} \bibnamefont{and}
  \bibinfo{author}{\bibfnamefont{M.}~\bibnamefont{Picco}},
  \bibinfo{journal}{Phys. Rev. E} \textbf{\bibinfo{volume}{61}},
  \bibinfo{pages}{R13} (\bibinfo{year}{2000}).

\bibitem[{\citenamefont{d'Auriac and Iglói}(2003)}]{AuIg03}
\bibinfo{author}{\bibfnamefont{J.~C.~A.} \bibnamefont{d'Auriac}}
  \bibnamefont{and} \bibinfo{author}{\bibfnamefont{F.}~\bibnamefont{Iglói}},
  \bibinfo{journal}{Phys. Rev. Lett.} \textbf{\bibinfo{volume}{90}},
  \bibinfo{pages}{190601} (\bibinfo{year}{2003}).

\bibitem[{\citenamefont{Muller-Krumbhaar and Stoll}(1976)}]{MuSt76}
\bibinfo{author}{\bibfnamefont{H.}~\bibnamefont{Muller-Krumbhaar}}
  \bibnamefont{and} \bibinfo{author}{\bibfnamefont{E.~P.} \bibnamefont{Stoll}},
  \bibinfo{journal}{J. Chem. Phys.} \textbf{\bibinfo{volume}{65}},
  \bibinfo{pages}{4294} (\bibinfo{year}{1976}).

\bibitem[{\citenamefont{Kunz and Souillard}(1978)}]{KuSo78a}
\bibinfo{author}{\bibfnamefont{H.}~\bibnamefont{Kunz}} \bibnamefont{and}
  \bibinfo{author}{\bibfnamefont{B.}~\bibnamefont{Souillard}},
  \bibinfo{journal}{Phys. Rev. Lett.} \textbf{\bibinfo{volume}{40}},
  \bibinfo{pages}{133} (\bibinfo{year}{1978}).

\bibitem[{\citenamefont{Schwartz}(1978)}]{Schwartz78}
\bibinfo{author}{\bibfnamefont{M.}~\bibnamefont{Schwartz}},
  \bibinfo{journal}{Phys. Rev. B} \textbf{\bibinfo{volume}{18}},
  \bibinfo{pages}{2364} (\bibinfo{year}{1978}).

\bibitem[{\citenamefont{Stoll and Domb}(1979)}]{StDo79}
\bibinfo{author}{\bibfnamefont{E.}~\bibnamefont{Stoll}} \bibnamefont{and}
  \bibinfo{author}{\bibfnamefont{C.}~\bibnamefont{Domb}}, \bibinfo{journal}{J.
  Phys. A: Math. Gen.} \textbf{\bibinfo{volume}{12}}, \bibinfo{pages}{1843}
  (\bibinfo{year}{1979}).

\bibitem[{\citenamefont{Hoshen et~al.}(1979)\citenamefont{Hoshen, Stauffer,
  Bishop, Harrison, and Quinn}}]{Hoshen79}
\bibinfo{author}{\bibfnamefont{H.}~\bibnamefont{Hoshen}},
  \bibinfo{author}{\bibfnamefont{D.}~\bibnamefont{Stauffer}},
  \bibinfo{author}{\bibfnamefont{G.~H.} \bibnamefont{Bishop}},
  \bibinfo{author}{\bibfnamefont{R.~J.} \bibnamefont{Harrison}},
  \bibnamefont{and} \bibinfo{author}{\bibfnamefont{G.~D.} \bibnamefont{Quinn}},
  \bibinfo{journal}{J. Phys. A: Math. Gen.} \textbf{\bibinfo{volume}{12}},
  \bibinfo{pages}{1285} (\bibinfo{year}{1979}).

\bibitem[{\citenamefont{Schmittmann and Bruce}(1985)}]{ScBr85}
\bibinfo{author}{\bibfnamefont{B.}~\bibnamefont{Schmittmann}} \bibnamefont{and}
  \bibinfo{author}{\bibfnamefont{A.~D.} \bibnamefont{Bruce}},
  \bibinfo{journal}{J. Phys. A} \textbf{\bibinfo{volume}{18}},
  \bibinfo{pages}{1715} (\bibinfo{year}{1985}).

\bibitem[{\citenamefont{Parisi and Sourlas}(1981)}]{PaSo81}
\bibinfo{author}{\bibfnamefont{G.}~\bibnamefont{Parisi}} \bibnamefont{and}
  \bibinfo{author}{\bibfnamefont{N.}~\bibnamefont{Sourlas}},
  \bibinfo{journal}{Phys. Rev. Lett.} \textbf{\bibinfo{volume}{46}},
  \bibinfo{pages}{871} (\bibinfo{year}{1981}).

\bibitem[{\citenamefont{Stauffer and Aharony}(1994)}]{StAh94}
\bibinfo{author}{\bibfnamefont{D.}~\bibnamefont{Stauffer}} \bibnamefont{and}
  \bibinfo{author}{\bibfnamefont{A.}~\bibnamefont{Aharony}},
  \emph{\bibinfo{title}{Introduction to Percolation Theory}}
  (\bibinfo{publisher}{Taylor \& Francis}, \bibinfo{address}{London},
  \bibinfo{year}{1994}).

\bibitem[{\citenamefont{Stella and Vanderzande}(1989)}]{StVa89}
\bibinfo{author}{\bibfnamefont{A.}~\bibnamefont{Stella}} \bibnamefont{and}
  \bibinfo{author}{\bibfnamefont{C.}~\bibnamefont{Vanderzande}},
  \bibinfo{journal}{Phys. Rev. Lett.} \textbf{\bibinfo{volume}{62}},
  \bibinfo{pages}{1067} (\bibinfo{year}{1989}).

\bibitem[{\citenamefont{Deroulers and Young}(2002)}]{DeYo02}
\bibinfo{author}{\bibfnamefont{C.}~\bibnamefont{Deroulers}} \bibnamefont{and}
  \bibinfo{author}{\bibfnamefont{A.~P.} \bibnamefont{Young}},
  \bibinfo{journal}{Phys. Rev. B} \textbf{\bibinfo{volume}{66}},
  \bibinfo{pages}{014438} (\bibinfo{year}{2002}).

\bibitem[{\citenamefont{Fradkov and Udler}(1994)}]{FrUd94}
\bibinfo{author}{\bibfnamefont{V.~E.} \bibnamefont{Fradkov}} \bibnamefont{and}
  \bibinfo{author}{\bibfnamefont{D.}~\bibnamefont{Udler}},
  \bibinfo{journal}{Adv. Phys.} \textbf{\bibinfo{volume}{43}},
  \bibinfo{pages}{739} (\bibinfo{year}{1994}).

\bibitem[{\citenamefont{Mullins}(1998)}]{Mullins98}
\bibinfo{author}{\bibfnamefont{W.~W.} \bibnamefont{Mullins}},
  \bibinfo{journal}{Acta mater.} \textbf{\bibinfo{volume}{46}},
  \bibinfo{pages}{6219} (\bibinfo{year}{1998}).

\bibitem[{\citenamefont{Rios and Lucke}(2001)}]{RiLu01}
\bibinfo{author}{\bibfnamefont{P.~R.} \bibnamefont{Rios}} \bibnamefont{and}
  \bibinfo{author}{\bibfnamefont{K.}~\bibnamefont{Lucke}},
  \bibinfo{journal}{Scripta Materialia} \textbf{\bibinfo{volume}{44}},
  \bibinfo{pages}{2471} (\bibinfo{year}{2001}).

\bibitem[{\citenamefont{Flyvbjerg}(1993)}]{Flyvbjerg93a}
\bibinfo{author}{\bibfnamefont{H.}~\bibnamefont{Flyvbjerg}},
  \bibinfo{journal}{Phys. Rev. E} \textbf{\bibinfo{volume}{47}},
  \bibinfo{pages}{4037} (\bibinfo{year}{1993}).

\bibitem[{\citenamefont{Hui and Berker}(1989)}]{HuBe89}
\bibinfo{author}{\bibfnamefont{K.}~\bibnamefont{Hui}} \bibnamefont{and}
  \bibinfo{author}{\bibfnamefont{A.~N.} \bibnamefont{Berker}},
  \bibinfo{journal}{Phys. Rev. Lett.} \textbf{\bibinfo{volume}{62}},
  \bibinfo{pages}{2507} (\bibinfo{year}{1989}).

\bibitem[{\citenamefont{Aizenman and Wehr}(1989)}]{AiWe89}
\bibinfo{author}{\bibfnamefont{M.}~\bibnamefont{Aizenman}} \bibnamefont{and}
  \bibinfo{author}{\bibfnamefont{J.}~\bibnamefont{Wehr}},
  \bibinfo{journal}{Phys. Rev. Lett.} \textbf{\bibinfo{volume}{62}},
  \bibinfo{pages}{2503} (\bibinfo{year}{1989}).

\bibitem[{\citenamefont{Chen et~al.}(1992)\citenamefont{Chen, Ferrenberg, and
  Landau}}]{ChFeLa92}
\bibinfo{author}{\bibfnamefont{S.}~\bibnamefont{Chen}},
  \bibinfo{author}{\bibfnamefont{A.~M.} \bibnamefont{Ferrenberg}},
  \bibnamefont{and} \bibinfo{author}{\bibfnamefont{D.~P.}
  \bibnamefont{Landau}}, \bibinfo{journal}{Phys. Rev. Lett.1}
  \textbf{\bibinfo{volume}{69}}, \bibinfo{pages}{1213} (\bibinfo{year}{1992}).

\bibitem[{\citenamefont{Ludwig}(1990)}]{Ludwig90}
\bibinfo{author}{\bibfnamefont{A.~W.~W.} \bibnamefont{Ludwig}},
  \bibinfo{journal}{Nuclear Physics B} \textbf{\bibinfo{volume}{330}},
  \bibinfo{pages}{639} (\bibinfo{year}{1990}).

\bibitem[{\citenamefont{Berche and Chatelain}(2004)}]{BeCh04}
\bibinfo{author}{\bibfnamefont{B.}~\bibnamefont{Berche}} \bibnamefont{and}
  \bibinfo{author}{\bibfnamefont{C.}~\bibnamefont{Chatelain}}, in
  \emph{\bibinfo{booktitle}{Order, disorder, and criticality}}, edited by
  \bibinfo{editor}{\bibfnamefont{Y.}~\bibnamefont{Holovatch}}
  (\bibinfo{publisher}{World Scientific}, \bibinfo{address}{Singapore},
  \bibinfo{year}{2004}), p. \bibinfo{pages}{146}.

\bibitem[{\citenamefont{Kinzel and Domany}(1981)}]{KiDo81}
\bibinfo{author}{\bibfnamefont{W.}~\bibnamefont{Kinzel}} \bibnamefont{and}
  \bibinfo{author}{\bibfnamefont{E.}~\bibnamefont{Domany}},
  \bibinfo{journal}{Phys. Rev. B} \textbf{\bibinfo{volume}{23}},
  \bibinfo{pages}{3421} (\bibinfo{year}{1981}).

\bibitem[{\citenamefont{Srolovitz and Grest}(1985)}]{SrGr85}
\bibinfo{author}{\bibfnamefont{D.~J.} \bibnamefont{Srolovitz}}
  \bibnamefont{and} \bibinfo{author}{\bibfnamefont{G.~S.} \bibnamefont{Grest}},
  \bibinfo{journal}{Phys. Rev. B} \textbf{\bibinfo{volume}{32}},
  \bibinfo{pages}{3021} (\bibinfo{year}{1985}).

\bibitem[{\citenamefont{Hazzledine and Oldershaw}(1990)}]{HaOl90}
\bibinfo{author}{\bibfnamefont{P.~M.} \bibnamefont{Hazzledine}}
  \bibnamefont{and} \bibinfo{author}{\bibfnamefont{R.~D.~J.}
  \bibnamefont{Oldershaw}}, \bibinfo{journal}{Philosophical Magazine A}
  \textbf{\bibinfo{volume}{61}}, \bibinfo{pages}{579} (\bibinfo{year}{1990}).

\bibitem[{\citenamefont{Krichevsky and Stavans}(1992)}]{KrSt92}
\bibinfo{author}{\bibfnamefont{O.}~\bibnamefont{Krichevsky}} \bibnamefont{and}
  \bibinfo{author}{\bibfnamefont{J.}~\bibnamefont{Stavans}},
  \bibinfo{journal}{Phys. Rev. B} \textbf{\bibinfo{volume}{46}},
  \bibinfo{pages}{10579} (\bibinfo{year}{1992}).

\bibitem[{\citenamefont{Paul et~al.}(2004)\citenamefont{Paul, Puri, and
  Rieger}}]{PaPuRi04}
\bibinfo{author}{\bibfnamefont{R.}~\bibnamefont{Paul}},
  \bibinfo{author}{\bibfnamefont{S.}~\bibnamefont{Puri}}, \bibnamefont{and}
  \bibinfo{author}{\bibfnamefont{H.}~\bibnamefont{Rieger}},
  \bibinfo{journal}{Europhys. Lett.} \textbf{\bibinfo{volume}{68}},
  \bibinfo{pages}{881} (\bibinfo{year}{2004}).

\bibitem[{\citenamefont{Paul et~al.}(2005)\citenamefont{Paul, Puri, and
  Rieger}}]{PaPuRi05}
\bibinfo{author}{\bibfnamefont{R.}~\bibnamefont{Paul}},
  \bibinfo{author}{\bibfnamefont{S.}~\bibnamefont{Puri}}, \bibnamefont{and}
  \bibinfo{author}{\bibfnamefont{H.}~\bibnamefont{Rieger}},
  \bibinfo{journal}{Phys. Rev. E} \textbf{\bibinfo{volume}{71}},
  \bibinfo{pages}{061109} (\bibinfo{year}{2005}).

\bibitem[{\citenamefont{Henkel and Pleimling}(2006)}]{HePl06}
\bibinfo{author}{\bibfnamefont{M.}~\bibnamefont{Henkel}} \bibnamefont{and}
  \bibinfo{author}{\bibfnamefont{M.}~\bibnamefont{Pleimling}},
  \bibinfo{journal}{Europhys. Lett.} \textbf{\bibinfo{volume}{76}},
  \bibinfo{pages}{561} (\bibinfo{year}{2006}).

\bibitem[{\citenamefont{Paul et~al.}(2007)\citenamefont{Paul, Schehr, and
  Rieger}}]{PaScRi07}
\bibinfo{author}{\bibfnamefont{R.}~\bibnamefont{Paul}},
  \bibinfo{author}{\bibfnamefont{G.}~\bibnamefont{Schehr}}, \bibnamefont{and}
  \bibinfo{author}{\bibfnamefont{H.}~\bibnamefont{Rieger}},
  \bibinfo{journal}{Phys. Rev. E} \textbf{\bibinfo{volume}{75}},
  \bibinfo{pages}{030104(R)} (\bibinfo{year}{2007}).

\bibitem[{\citenamefont{Iguain et~al.}(2009)\citenamefont{Iguain, Bustingorry,
  Kolton, and Cugliandolo}}]{Igbukocu09}
\bibinfo{author}{\bibfnamefont{J.~L.} \bibnamefont{Iguain}},
  \bibinfo{author}{\bibfnamefont{S.}~\bibnamefont{Bustingorry}},
  \bibinfo{author}{\bibfnamefont{A.~B.} \bibnamefont{Kolton}},
  \bibnamefont{and} \bibinfo{author}{\bibfnamefont{L.~F.}
  \bibnamefont{Cugliandolo}}, \bibinfo{journal}{Phys. Rev. B}
  \textbf{\bibinfo{volume}{80}}, \bibinfo{pages}{094201}
  (\bibinfo{year}{2009}).

\bibitem[{\citenamefont{Sicilia
  et~al.}(2008{\natexlab{a}})\citenamefont{Sicilia, Arenzon, Bray, and
  Cugliandolo}}]{SiArBrCu08}
\bibinfo{author}{\bibfnamefont{A.}~\bibnamefont{Sicilia}},
  \bibinfo{author}{\bibfnamefont{J.~J.} \bibnamefont{Arenzon}},
  \bibinfo{author}{\bibfnamefont{A.~J.} \bibnamefont{Bray}}, \bibnamefont{and}
  \bibinfo{author}{\bibfnamefont{L.~F.} \bibnamefont{Cugliandolo}},
  \bibinfo{journal}{Europhys. Lett.} \textbf{\bibinfo{volume}{82}},
  \bibinfo{pages}{10001} (\bibinfo{year}{2008}{\natexlab{a}}).

\bibitem[{\citenamefont{Picco}(1998)}]{Picco98}
\bibinfo{author}{\bibfnamefont{M.}~\bibnamefont{Picco}} (\bibinfo{year}{1998}),
  \bibinfo{note}{cond-mat/9802092}.

\bibitem[{\citenamefont{Olson and Young}(1999)}]{OlYo99}
\bibinfo{author}{\bibfnamefont{T.}~\bibnamefont{Olson}} \bibnamefont{and}
  \bibinfo{author}{\bibfnamefont{A.~P.} \bibnamefont{Young}},
  \bibinfo{journal}{Phys. Rev. B} \textbf{\bibinfo{volume}{60}},
  \bibinfo{pages}{3428} (\bibinfo{year}{1999}).

\bibitem[{\citenamefont{Jacobsen and Cardy}(1998)}]{JaCa98}
\bibinfo{author}{\bibfnamefont{J.~L.} \bibnamefont{Jacobsen}} \bibnamefont{and}
  \bibinfo{author}{\bibfnamefont{J.}~\bibnamefont{Cardy}},
  \bibinfo{journal}{Nuc. Phys. B} \textbf{\bibinfo{volume}{515}},
  \bibinfo{pages}{701} (\bibinfo{year}{1998}).

\bibitem[{\citenamefont{Schliecker}(2002)}]{Schliecker02}
\bibinfo{author}{\bibfnamefont{G.}~\bibnamefont{Schliecker}},
  \bibinfo{journal}{Adv. Phys.} \textbf{\bibinfo{volume}{51}},
  \bibinfo{pages}{1319} (\bibinfo{year}{2002}).

\bibitem[{\citenamefont{Sicilia
  et~al.}(2008{\natexlab{b}})\citenamefont{Sicilia, Arenzon, Dierking, Bray,
  Cugliandolo, Martinez-Perdiguero, Alonso, and Pintre}}]{Si08}
\bibinfo{author}{\bibfnamefont{A.}~\bibnamefont{Sicilia}},
  \bibinfo{author}{\bibfnamefont{J.~J.} \bibnamefont{Arenzon}},
  \bibinfo{author}{\bibfnamefont{I.}~\bibnamefont{Dierking}},
  \bibinfo{author}{\bibfnamefont{A.~J.} \bibnamefont{Bray}},
  \bibinfo{author}{\bibfnamefont{L.~F.} \bibnamefont{Cugliandolo}},
  \bibinfo{author}{\bibfnamefont{J.}~\bibnamefont{Martinez-Perdiguero}},
  \bibinfo{author}{\bibfnamefont{I.}~\bibnamefont{Alonso}}, \bibnamefont{and}
  \bibinfo{author}{\bibfnamefont{I.~C.} \bibnamefont{Pintre}},
  \bibinfo{journal}{Phys. Rev. Lett.} \textbf{\bibinfo{volume}{101}},
  \bibinfo{pages}{197801} (\bibinfo{year}{2008}{\natexlab{b}}).

\end{thebibliography}

\end{document}